\documentclass[preprint2,twocolappendix]{aastex631}

\shorttitle{JWST NIRISS - IV AMI}
\shortauthors{Sivaramakrishnan et al.}

\usepackage{xcolor}
\definecolor{nIRpink}{RGB}{255,240,240}
\usepackage{graphicx}
\usepackage{bm}
\usepackage{booktabs}
\usepackage{float}

\newcommand{\ie}{\textit{i}.\textit{e}.}
\newcommand{\eg}{\textit{e}.\textit{g}.}

\newcommand{\nicemicron}{$\text{\textmu m}$}

\newcommand{\uvplane}{$uv$-plane}
\newcommand{\mirage}{\texttt{MIRaGe}}

\begin{document}

\title{The Near Infrared Imager and Slitless Spectrograph for 
    the  James Webb Space Telescope - 
    IV. Aperture Masking Interferometry}

\correspondingauthor{Anand Sivaramakrishnan}
\email{anand@stsci.edu}

\author[0000-0003-1251-4124]{Anand Sivaramakrishnan}
\affiliation{Space Telescope Science Institute, 3700 San Martin Drive, Baltimore, MD 21218, USA}
\affiliation{Astrophysics Department, American Museum of Natural History, 79th Street at Central Park West, New York, NY 10024}
\affiliation{Department of Physics and Astronomy, Johns Hopkins University, 3701 San Martin Drive, Baltimore, MD 21218, USA}

\author[0000-0001-7026-6291]{Peter Tuthill}
\affiliation{School of Physics, University of Sydney, NSW 2006, Australia}

\author{James P. Lloyd}
\affiliation{Department of Space Physics, Cornell University, ... Ithaca NY}

\author[0000-0002-7162-8036]{Alexandra Z. Greenbaum}
\affiliation{IPAC, Mail Code 100-22, Caltech, 1200 E. California Blvd., Pasadena, CA 91125, USA}

\author{Deepashri Thatte}
\affiliation{Space Telescope Science Institute, 3700 San Martin Drive, Baltimore, MD 21218, USA}

\author[0000-0001-7864-308X]{Rachel A. Cooper}
\affiliation{Space Telescope Science Institute, 3700 San Martin Drive, Baltimore, MD 21218, USA}

\author[0000-0002-5922-8267]{Thomas Vandal}
\affiliation{Institut de Recherche sur les Exoplan\`etes (iREx), Universit\'e de Montr\'eal, D\'epartement de Physique, \\ C.P. 6128 Succ. Centre-ville, Montr\'eal,  QC H3C 3J7, Canada.}

\author[0000-0003-2769-0438]{Jens Kammerer}
\affiliation{Space Telescope Science Institute, 3700 San Martin Drive, Baltimore, MD 21218, USA}

\author[0000-0002-9723-0421]{Joel Sanchez-Bermudez}
\affiliation{Instituto de Astronom\'ia, Universidad Nacional Aut\'onoma de M\'exico, Apdo. Postal 70264, Ciudad de M\'exico, 04510, M\'exico}
\affiliation{Max-Planck-Institut f\"ur Astronomie, K\"onigstuhl 17, D-69117 Heidelberg, Germany}

\author[0000-0003-2595-9114]{Benjamin J. S. Pope}
\affiliation{School of Mathematics and Physics, The University of Queensland, St Lucia, QLD 4072, Australia}
\affiliation{Centre for Astrophysics, University of Southern Queensland, West Street, Toowoomba, QLD 4350, Australia}

\author[0000-0001-9582-4261]{Dori Blakely}
\affiliation{Department of Physics and Astronomy, University of Victoria, 3800 Finnerty Road, Elliot Building, Victoria, BC, V8P 5C2, Canada}
\affiliation{NRC Herzberg Astronomy and Astrophysics, 5071 West Saanich Road, Victoria, BC, V9E 2E7, Canada}

\author[0000-0003-0475-9375]{Lo\"ic Albert}
\affiliation{Institut de Recherche sur les Exoplan\`etes (iREx), Universit\'e de Montr\'eal, D\'epartement de Physique, \\ C.P. 6128 Succ. Centre-ville, Montr\'eal,  QC H3C 3J7, Canada.}

\author[0000-0003-4166-4121]{Neil J. Cook} 
\affiliation{Institut de Recherche sur les Exoplan\`etes (iREx), Universit\'e de Montr\'eal, D\'epartement de Physique, \\ C.P. 6128 Succ. Centre-ville, Montr\'eal,  QC H3C 3J7, Canada.} 

\author[0000-0002-6773-459X]{Doug Johnstone}
\affiliation{NRC Herzberg Astronomy and Astrophysics, 5071 West Saanich Road, Victoria, BC, V9E 2E7, Canada}
\affiliation{Department of Physics and Astronomy, University of Victoria, 3800 Finnerty Road, Elliot Building, Victoria, BC, V8P 5C2, Canada}

\author{Andr\'e R. Martel}
\affiliation{Space Telescope Science Institute, 3700 San Martin Drive, Baltimore, MD 21218, USA}

\author{Kevin Volk}
\affiliation{Space Telescope Science Institute, 3700 San Martin Drive, Baltimore, MD 21218, USA}

\author[0000-0001-7661-5130]{Anthony Soulain}
\affiliation{Univ. Grenoble Alpes, CNRS, IPAG }

\author[0000-0003-3506-5667]{\'Etienne Artigau}
\affiliation{Institut de Recherche sur les Exoplan\`etes (iREx), Universit\'e de Montr\'eal, D\'epartement de Physique, \\ C.P. 6128 Succ. Centre-ville, Montr\'eal,  QC H3C 3J7, Canada.}

\author[0000-0002-6780-4252]{David Lafreni\`ere}
\affiliation{Institut de Recherche sur les Exoplan\`etes (iREx), Universit\'e de Montr\'eal, D\'epartement de Physique, \\ C.P. 6128 Succ. Centre-ville, Montr\'eal,  QC H3C 3J7, Canada.}

\author[0000-0002-4201-7367]{Chris J. Willott}
\affil{NRC Herzberg, 5071 West Saanich Rd, Victoria, BC V9E 2E7, Canada}

\author{S\'ebastien Parmentier}
\affiliation{Grafham Grange School, Grafham, Bramley, Guildford GU5 0LH, UK}

\author[0000-0002-5956-851X]{K. E. Saavik Ford}
\affiliation{ Department of Science, Borough of Manhattan Community College, City University of New York, New York, NY 10007, USA} 
\affiliation{Astrophysics Department, American Museum of Natural History, 79th Street at Central Park West, New York, NY 10024}

\author[0000-0002-9726-0508]{Barry McKernan}
\affiliation{ Department of Science, Borough of Manhattan Community College, City University of New York, New York, NY 10007, USA} 
\affiliation{Astrophysics Department, American Museum of Natural History, 79th Street at Central Park West, New York, NY 10024}

\author[0000-0003-3504-1569]{M. Bego\~na Vila}
\affiliation{NASA Goddard Space Flight Center, 8800 Greenbelt Rd, Greenbelt, MD 20771}
\affiliation{KBR Space Engineering Division, 8120 Maple Lawn Blvd, Fulton, MD 20759}

\author[0000-0002-1715-7069]{Neil Rowlands}
\affiliation{Honeywell Aerospace \#100, 303 Terry Fox Drive, Ottawa,  ON  K2K 3J1, Canada } 

\author[0000-0001-5485-4675]{Ren\'e Doyon}
\affiliation{Institut de Recherche sur les Exoplan\`etes (iREx), Universit\'e de Montr\'eal, D\'epartement de Physique, \\ C.P. 6128 Succ. Centre-ville, Montr\'eal,  QC H3C 3J7, Canada.}


\author{ Louis Desdoigts}
\affiliation{School of Physics, University of Sydney, NSW 2006, Australia}

\author[0000-0003-2429-7964]{Alexander W. Fullerton}
\affiliation{Space Telescope Science Institute, 3700 San Martin Drive, Baltimore, MD 21218, USA}

\author[0000-0003-1863-4960]{Matthew De Furio}
\affiliation{University of Michigan, Ann Arbor, MI}

\author[0000-0002-5728-1427]{Paul Goudfrooij}
\affiliation{Space Telescope Science Institute, 3700 San Martin Drive, Baltimore, MD 21218, USA}

\author[0000-0002-7092-2022]{Sherie T. Holfeltz}
\affiliation{Space Telescope Science Institute, 3700 San Martin Drive, Baltimore, MD 21218, USA}

\author[0000-0002-5907-3330]{Stephanie LaMassa}
\affiliation{Space Telescope Science Institute, 3700 San Martin Drive, Baltimore, MD 21218, USA}

\author{Michael Maszkiewicz}
\affiliation{Canadian Space Agency, 6767 route de l'Aéroport, Saint-Hubert, Quebec, J3V 8Y9, Canada}

\author[0000-0003-1227-3084]{Michael R. Meyer}
\affiliation{University of Michigan, Ann Arbor, MI}

\author[0000-0002-3191-8151]{Marshall D. Perrin}
\affiliation{Space Telescope Science Institute, 3700 San Martin Drive, Baltimore, MD 21218, USA}

\author[0000-0003-3818-408X]{Laurent Pueyo}
\affiliation{Space Telescope Science Institute, 3700 San Martin Drive, Baltimore, MD 21218, USA}

\author[0000-0001-9525-3673]{Johannes Sahlmann}
\affiliation{RHEA Group for the European Space Agency (ESA), European Space Astronomy Centre (ESAC),\\ Camino Bajo del Castillo s/n, 28692 Villanueva de la Ca\~nada, Madrid, Spain}

\author[0000-0001-8368-0221]{Sangmo Tony Sohn}
\affiliation{Space Telescope Science Institute, 3700 San Martin Drive, Baltimore, MD 21218, USA}

\author[0000-0002-3665-5784]{Paula S. Teixeira}
\affiliation{ Scottish Universities Physics Alliance (SUPA), School of Physics and Astronomy, University of St Andrews, North Haugh, Fife, KY16 9SS, St Andrews, UK}


\begin{abstract}
The James Webb Space Telescope's Near Infrared Imager and Slitless Spectrograph (JWST-NIRISS) flies a 7-hole non-redundant mask (NRM), the first such interferometer in space, operating at 3-5 \nicemicron~wavelengths, and a bright limit of $\simeq 4$ magnitudes in W2.   We describe the NIRISS Aperture Masking Interferometry (AMI) mode to help potential observers understand its underlying principles, present some sample science cases, explain its operational observing strategies, indicate how AMI proposals can be developed with data simulations, and how AMI data can be analyzed.  We also present key results from commissioning AMI.  Since the allied Kernel Phase Imaging (KPI) technique benefits from AMI operational strategies, we also cover NIRISS KPI methods and analysis techniques, including a new user-friendly KPI pipeline.  The NIRISS KPI bright limit is $\simeq 8$ W2 magnitudes.  AMI (and KPI) achieve an inner working angle of $\sim  70$ mas that is well inside the $\sim 400$ mas NIRCam inner working angle for its circular occulter coronagraphs at comparable wavelengths.

\end{abstract}


\section{Introduction}
\label{sec:Intro}

After its successful launch and commissioning, the James Webb Space Telescope (JWST) is already beginning to open new windows on the universe. Its four scientific instruments operating in the near-infrared (near-IR) and mid-infrared (mid-IR) provide multiple observing modes designed to probe a wide range of celestial phenomena, from structure formation in the early universe to the birth of extrasolar planets around nearby stars, as well as objects within our own solar system.

Three of JWST's four instruments offer high contrast imaging (HCI) modes that enable imaging of faint companions and faint extended emission around bright sources.  Typical targets include exoplanets, circumstellar environments, and active galaxy nuclei (AGN). High contrast is obtained by minimizing the deleterious impact of light from the bright primary target by dint of a combination of optical, observational, and post-processing techniques, revealing the presence of nearby faint companions or structures. HCI with JWST uses either coronagraphy or interferometry. The Near-Infrared Camera \citep[NIRCam,][]{2009SPIE.7440E..0WK} and the Mid-Infrared Instrument \citep[MIRI,][]{2015PASP..127..633B} provide coronagraphic imaging at 2-5\,\nicemicron~and 10-23\,\nicemicron, respectively. While NIRCam and MIRI coronagraphy have demonstrated a contrast of $\sim15$~mag and $\sim12$~mag, respectively, at wide separations \citep{2022arXiv220800998G,boccaletti2022arXiv220711080B,carter2022arXiv220814990C}, their performance is significantly reduced close to their inner working angle (IWA) of 0.4\arcsec and 0.3\arcsec.\footnote{
    \href{https://jwst-docs.stsci.edu/jwst-mid-infrared-instrument/miri-observing-modes/miri-coronagraphic-imaging}{MIRI coronagraphic imaging}}


The Near-Infrared Imager and Slitless Spectrograph (NIRISS, \citealt{2012SPIE.8442E..2RD}) has a unique HCI mode that is highly complementary to coronagraphy: Aperture Masking Interferometry \citep[AMI,][]{2009astro2010T..40S,2010SPIE.7731E..3WS,2012SPIE.8442E..2SS,2014A&A...567A..21S,2015ApJ...798...68G,2015SPIE.9605E..2FT,2016LPI....47.3005T,2016PASP..128a8006K,soulain_james_2020,2020SPIE11446E..1OS,2022SPIE_Tuthill}. 
The non-redundant mask (NRM) -- also called a sparse aperture mask (SAM) -- implements AMI with a 7-hole pupil mask in NIRISS. Among the earliest forms of interferometer devised \citep{fizeau1868,stephane1874,Michelson1891Natur,Michelson1921ApJ}, aperture masking returned to favor with astronomers in the last few decades as a technique for recovering information on the scale of the telescope's diffraction limit \citep{1995MNRAS.276..640H,2000PASP_Masking}.
With higher order wavefront control becoming ubiquitous at large ground-based observatories, masking found further application in recovery of high contrast companions at similar angular scales \citep{2006ApJ...650L.131L,SparseAO2006SPIE,Lacour2011A&A,LkCa152012ApJ,Sallum2015}. Exploiting the strengths of the  NRM, NIRISS AMI offers higher contrast science than images obtained with the full JWST aperture for the discovery space that lies very close to a bright PSF core. 
AMI's small IWA  is particularly relevant to imaging circumstellar environments at inner solar system scales around nearby stars, which can place stringent demands on small spatial scales.  Several well-explored extreme adaptive optics (ExAO) targets fall within AMI's reach, and observations of some brighter planetary systems in the near-solar environment may yield well resolved structures.

NIRISS AMI's specialized observing strategy is well-suited to a newer yet proven technique of Kernel Phase Imaging (KPI) analysis \citep{2010ApJ...724..464M,2013PASP..125..422M}, which has its origins in an AMI-style Fizeau approach to full aperture data.  On NIRISS, KPI accesses fainter targets than AMI while achieving similar IWAs.

The European Southern Observatory (ESO) MATISSE beam-combiner on the Very Large Telescope Interferometer  \citep[VLTI,][]{2022A&A...659A.192L} probes a spectral range which overlaps that of NIRISS AMI. Its $\sim\,$30 -- 130\,meter baselines afford MATISSE better angular resolution (3\,mas\,$< \theta <$\,13\,mas at $\lambda_0 = 3.8$~\nicemicron), whereas AMI's wider search space covers 70\,mas\,$ < \theta <$\,217\,mas. Used in tandem, AMI and MATISSE enable interferometry spanning scales between a few to a few hundred millarcseconds..

MATISSE is limited by thermal background: its $L$-band faint limit is 0.1 to 1\,Jy (in its low-spectral resolution mode). AMI's expected thermal backgrounds in F430M and F480M filters are  $\sim$0.4 and $\sim$0.5~electrons\,s$^{-1}\,\mathrm{pixel}^{-1}$ \citep{2015ApJ...798...68G}, just under 2~MJy\,s$^{-1}$; a substantial gain for targets where VLTI-MATISSE is background-limited. Similarly, ESO's new ERIS instrument \citep{eris2018SPIE10702E..09D} offers SAM in the L- and M-bands albeit at reduced sensitivity compared to JWST-NIRISS.

Combining AMI with VLTI-MATISSE should advance the study of  Active Galactic Nuclei (AGN). To date, VLTI-MATISSE has only been able to observe Circinus \citep{2022A&A...663A..35I} and NGC 1068 \citep{2022Natur.602..403G} because other sources are too faint. AMI is expected to provide unique constraints on the morphology of AGN at scales of a few hundred mas, contributing data to further theoretical models and engender subsequent observations with improved capabilities on ground-based beam-combiners.  

Other ground-based facilities such as VLTI-GRAVITY operating in the K-Band \citep{2017A&A...602A..94G}, aperture masking in the near-IR with adaptive optics systems SPHERE and the Gemini Planetary Imager (GPI) offer complementary near-IR coverage to add to the longer wavelength IR coverage of NIRISS AMI.

This paper develops its exposition of NIRISS AMI beginning with Section \ref{sec:nrmkpmath} where we define basic interferometric observables with focus on AMI and KPI.   
Four example AMI science cases are then outlined in Section \ref{sec:scienceexamples}. 
Section \ref{sec:ops} sketches the operational aspects of the NIRISS AMI mode, 
Section \ref{sec:analysis} the data analysis, and
\ref{sec:prep4prop} describes the mode's proposal preparation. Key results from commissioning NIRISS are highlighted in Section \ref{sec:perf}.
The appendix contains relevant details of the instrument.


\begin{figure*}[ht!]
  \centering
  \includegraphics[width=0.8\linewidth]{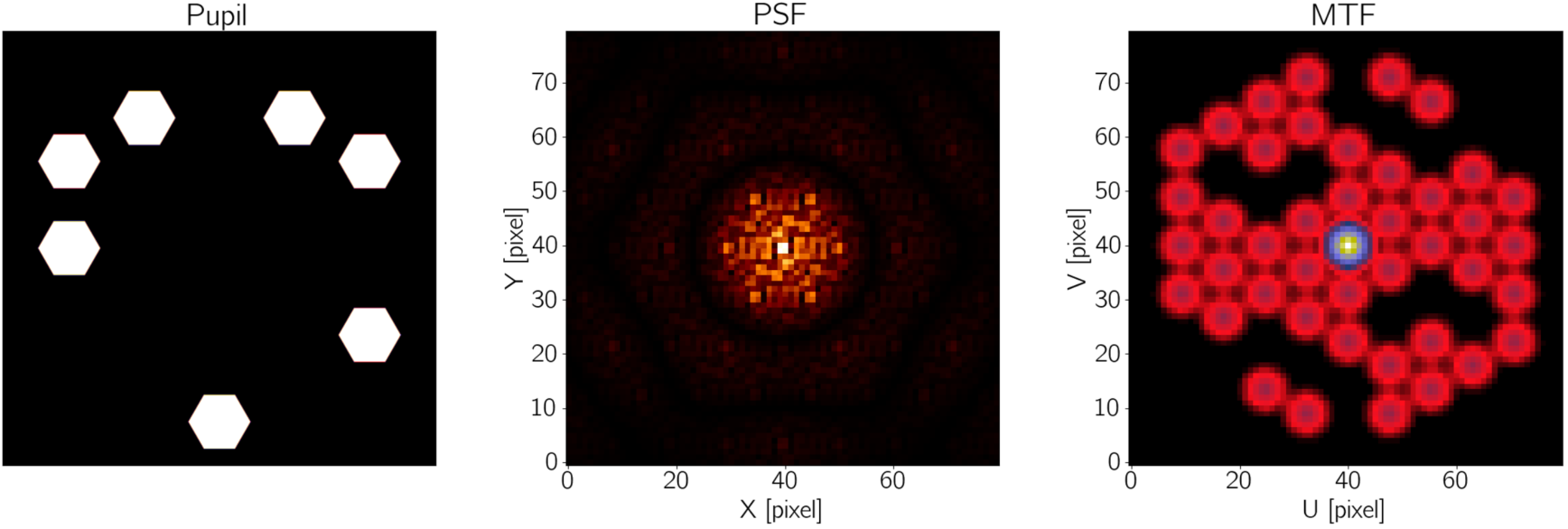} 
  \caption{The JWST-NIRISS AMI NRM, PSF, and its uv-coverage. \textbf{Left}: AMI's 7 hole pupil mask. \textbf{Middle}: a point source image taken through the NRM, showing a bright peak at the phase center, and with the primary beam crisscrossed by fringe patterns at 21 spatial frequencies. \textbf{Right}: The Fourier amplitude of the PSF. Note 21 ``splodges'' isolating regions of strong fringe signal (in red) and their mirrored Hermitian counterparts in this \uvplane~coverage, or Modulation Transfer Function (MTF), plot.} 
  \label{fig:pup_psf_mtf}
\end{figure*}

\section{Interferometry on JWST}
\label{sec:nrmkpmath}

We briefly explain the structure of the point-spread function (PSF) of a non-redundant mask (NRM), and outline the principles of JWST's Fizeau interferometry that underlie both AMI and KPI.  
Figure \ref{fig:pup_psf_mtf} shows the AMI NRM pupil, its PSF, and the absolute value of the PSF's Fourier transform in the \uvplane.  The latter is the Modulation Transfer Function (MTF) of the optical system.
A step-by-step account of NRM image formation can be found in \eg~ \citet{2015ApJ...798...68G}, and details of KPI imaging in \citet{2020A&A...636A..72M}.
Instrument details pertinent to AMI and KPI with NIRISS are presented in the appendix.

\subsection{Image formation with an NRM}
\label{ssec:nrm_math}


\newcommand \lap {$\stackrel{<}{\sim}$}
\newcommand \gap {$\stackrel{>}{\sim}$}

\newcommand \conv {{ * }}

\newcommand \Dlam 	{D_{\lambda}}
\newcommand \bul 	{{{\bf x}_\lambda}}
\newcommand \bu 	{{{\bf x}}}
\newcommand \ba 	{{\bf a}}
\newcommand \bx 	{{\bf u}}
\newcommand \br 	{{\bf r}}
\newcommand \bkl 	{{\bf k_{\lambda}}}
\newcommand \bk 	{{\bf k}}
\newcommand \bbmn 	{{\bf b}_{mn}}

\newcommand \CV   {$\mathcal{V}$}
\newcommand \ACF  {$\mathcal{ACF}$}
\newcommand \acf  {\mathcal{ACF}}
\newcommand \CCF  {$\mathcal{CCF}$}
\newcommand \ccf  {\mathcal{CCF}}
\newcommand \Sp     {\sum_{n=1}^N  \delta(\bu - \bu_{n})}
\newcommand \Spq     {\sum_{n,m} \delta(\bu_n - \bu_m)}
\newcommand \Epq     {\sum_{n,m} e^{-i\bk \cdot (\bu_n - \bu_m)}}
\newcommand \Ep     {\sum_{n} e^{-i\bk \cdot \bu_n }}

\newcommand \SHp     {\sum_{n}  \Big{(} H_n(\bu) e^{i \varphi_n(\bu)} \Big{)}  \conv \delta(\bu - \bu_n)}
\newcommand \SHP     {\sum_{n}    H_o(\bu)                           \conv \delta(\bu - \bu_n)  e^{i \phi_{n} }}

\newcommand \Nh     {N_{{\rm holes} }}
\
\newcommand \cv     {\mathcal{V}}

We describe a telescopic image of an infinitely distant on-axis source that emits spatially incoherent monochromatic radiation using the scalar wave approximation, in the Fraunhofer diffraction regime. The image plane complex amplitude from a point source at infinity is the 
Fourier transform\footnote{The FT of $A(\bu)$ is $a(\bk) \equiv \int_{-\infty}^{\infty} A(\bu) e^{-i\bk \cdot \bu} d\bu$}  (FT) of the complex aperture illumination function generated by the source \citep[\eg,][]{1999prop.book.....B}.
Position in the aperture (or pupil) plane is $\bul = (x/\lambda, y/\lambda)$, where $\lambda$ is the wavelength of the light under consideration.   Henceforth we drop the $\lambda$ suffix, making $\bu = (x,y)$ dimensionless.
A polychromatic image can be constructed as a sum or integral of appropriately weighted monochromatic intensity images.  

The NRM is applied to window the beam at a re-imaged aperture plane. We project back through the optics to the plane where incoming light is first restricted by the NRM's (projected) boundaries, since angular resolution is set in this plane.
The NRM has $N$ identical holes (or \textit{subapertures})  with centers $\{\bu_n, n=1,...,N\}$  and \textit{baselines} $\{\bbmn = \bu_n - \bu_m, n,m \in 1,...,N\}$.   No baseline in this set is repeated, so there are $N(N-1)/2$ independent baselines.  Hereafter we assume subscripts $m$ and $n$ run over subaperture indices $\{1,...,N\}$.

Position in the  image plane, $\bk = (k_x, k_y)$ has dimensions of  radians on the sky. Aberrations described by real-valued functions for subaperture-specific transmissions $\{ H_n(\bu) \}$ and phases $\{ \varphi_n(\bu) \}$ describe the wave in the aperture:
	\begin{equation} \label{ab}
		A(\bu) =  \SHp
	\end{equation}
(where $\conv$ denotes convolution). Assuming identical subaperture transmissions $H_n(\bu) = H_o(\bu)$, and the only phase errors being segment pistons $\{\phi_{n}\}$, with $\Delta_{mn} \equiv \phi_{m} - \phi_{n}$, the aperture complex amplitude is
	\begin{equation} \label{aa}
		A(\bu) =  \SHP,
	\end{equation}
with a PSF $p = a a^*$ (where $A$ and $a$ are a Fourier transform pair, denoted $A \rightleftharpoons a$):
    \begin{eqnarray}  \label{ac}
   p(\mathbf{k}) =&&P(\mathbf{k}) \times \Big\{N + \nonumber  \\
  \sum_{m>n}&2\big(&\cos{( \mathbf{k} \cdot \bbmn)}  \cos{\Delta_{mn})}  \nonumber  \\ 
                  &-&\sin{( \mathbf{k} \cdot \bbmn)}  \sin{(\Delta_{mn})}\big)\Big\}.
    \end{eqnarray}
$\Delta_{mn}$ is the fringe phase associated with baseline $\bbmn$; in a PSF it is the piston difference between the subapertures forming the baseline. The \textit{primary beam} $P(\bk) \equiv h_o \conv h^*_o$ is a single hole's PSF.
The primary beam's core is the interferometer's field of view.  The image  of an arbitrary source that does not fill the primary beam is well-described by the interferogram model in Equation (\ref{ac}) using its baselines and $N(N-1)/2$ constants $\{(a_{mn}, b_{mn})\}$  in lieu of $\{\cos(\Delta_{mn})\},\{\sin(\Delta_{mn})\}$.
The image plane intensity is traversed by $N(N-1)/2$ \textit{intensity} fringes, one per baseline. Each fringe is decentered by its \textbf{fringe} (or \textbf{Fourier}) \textbf{phase} by an angular fringe shift  of
	\begin{equation} \label{ad}
    \Delta\phi_{mn} = \mathrm{arctan}(b_{mn}/a_{mn})
	\end{equation}
and a \textbf{squared (fringe) visibility}
    \begin{equation} 
    \label{ad2}
    b_{mn}^2 + a_{mn}^2 = \cv_{mn} \cv^*_{mn} = V_{mn}^2.
    \end{equation}
\textbf{Raw interferometric observables}---fringe phases and fringe visibility amplitudes---can be derived from image data by fitting the image with the PSF model in Equation (\ref{ac}) as long as an analytical or precomputed numerical function $h_o$ is known at each pixel center:
	\begin{equation} \label{ae}
        \mathrm{Image\  data} = F~p(\mathbf{k}) + C.
    \end{equation}
The linear fitting problem yields $F$ (where $NF$ is the total flux),  $\{a_{mn}\}$ and $\{b_{mn}\}$ which provides $\{\Delta\phi_{mn}\}$ using Equation (\ref{ad}) and $\{V_{mn}\}$ using Equation (\ref{ad2}), and $C$, a constant pedestal or ``DC offset'' in the image.    Complex visibilities $\{\cv_{mn} = V_{mn}~e^{i\Delta\phi_{mn}}\}$ contain source geometry and segment piston information.

A numerical approach for extracting observables also exists. The \textbf{complex visibility} $\cv(\bx)$ is the Fourier transform of an image intensity. Its value at the origin is the total power in the image.  We use \uvplane~$\bx$ coordinates with the same dimensionality as those of the aperture plane,  length measured in wavelengths. $\cv(\bx)$ is Hermitian since the image intensity is real.

The support of signal in the \uvplane~is twice the size of the aperture, as the former is the autocorrelation of the latter. The length of a baseline (in wavelengths), $|\bx|$, is the inverse of the angular resolution (in radians) provided by the baseline.  
The complex numbers at the location corresponding to the baselines in the numerical Fourier Transform of the image provide the complex visibilities:  $\cv(\bbmn) = V_{mn} e^{i \Delta\phi_{mn}}$.

In total, the NRM has $N(N-1)/2$ baselines, each of which possesses a fringe phase and fringe amplitude,
$N(N-1)(N-2)/6$ closure phases, 
$(N-1)(N-2)/2$ independent closure phases,
$N(N-1)(N-2)(N-3)/(4!)$ closure amplitudes,
and
$N(N-3)/2$ independent closure amplitudes. 


\begin{figure*}[ht!]
\centering
\includegraphics[scale=0.425]{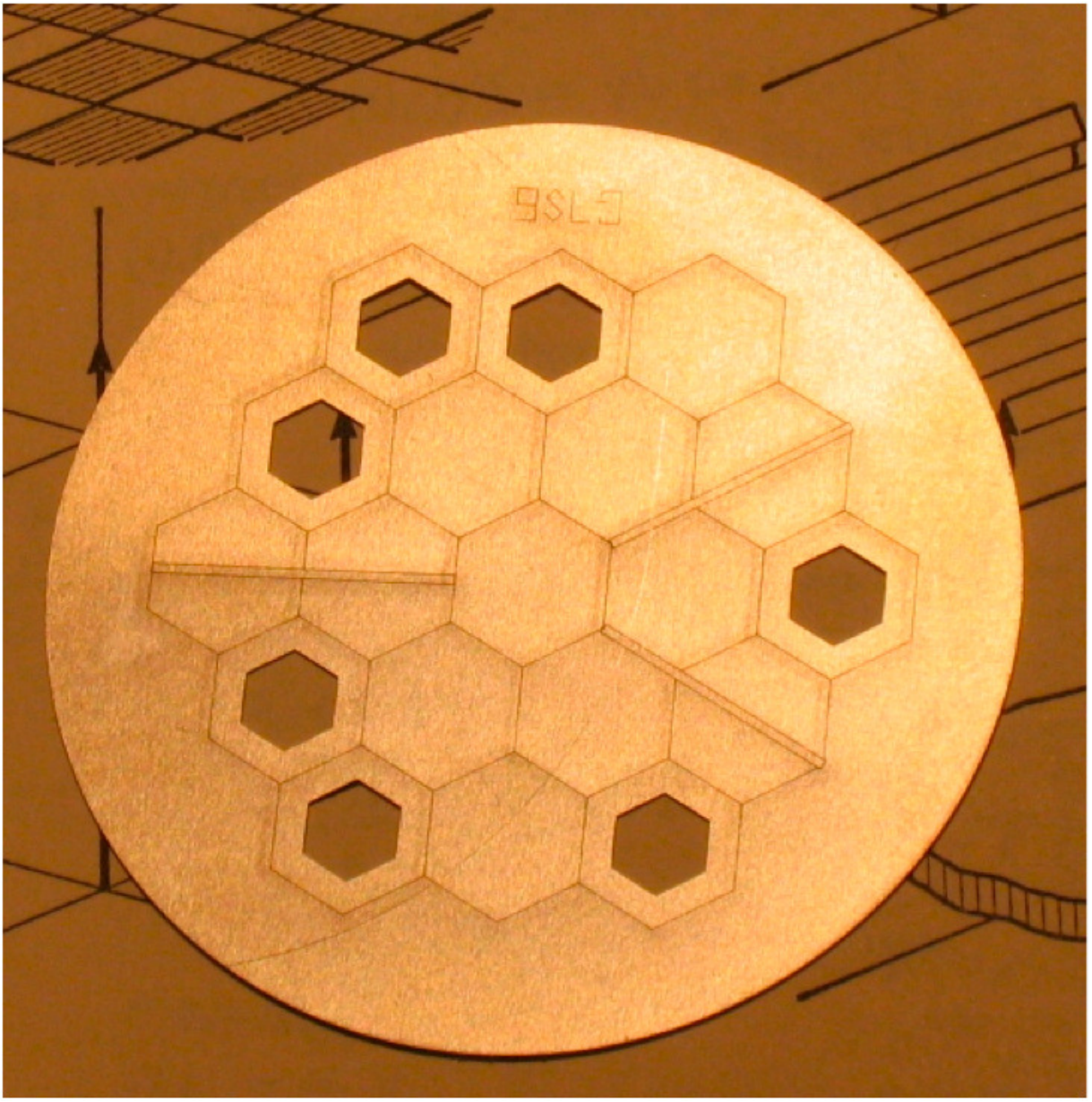} \quad
\includegraphics[scale=0.3]{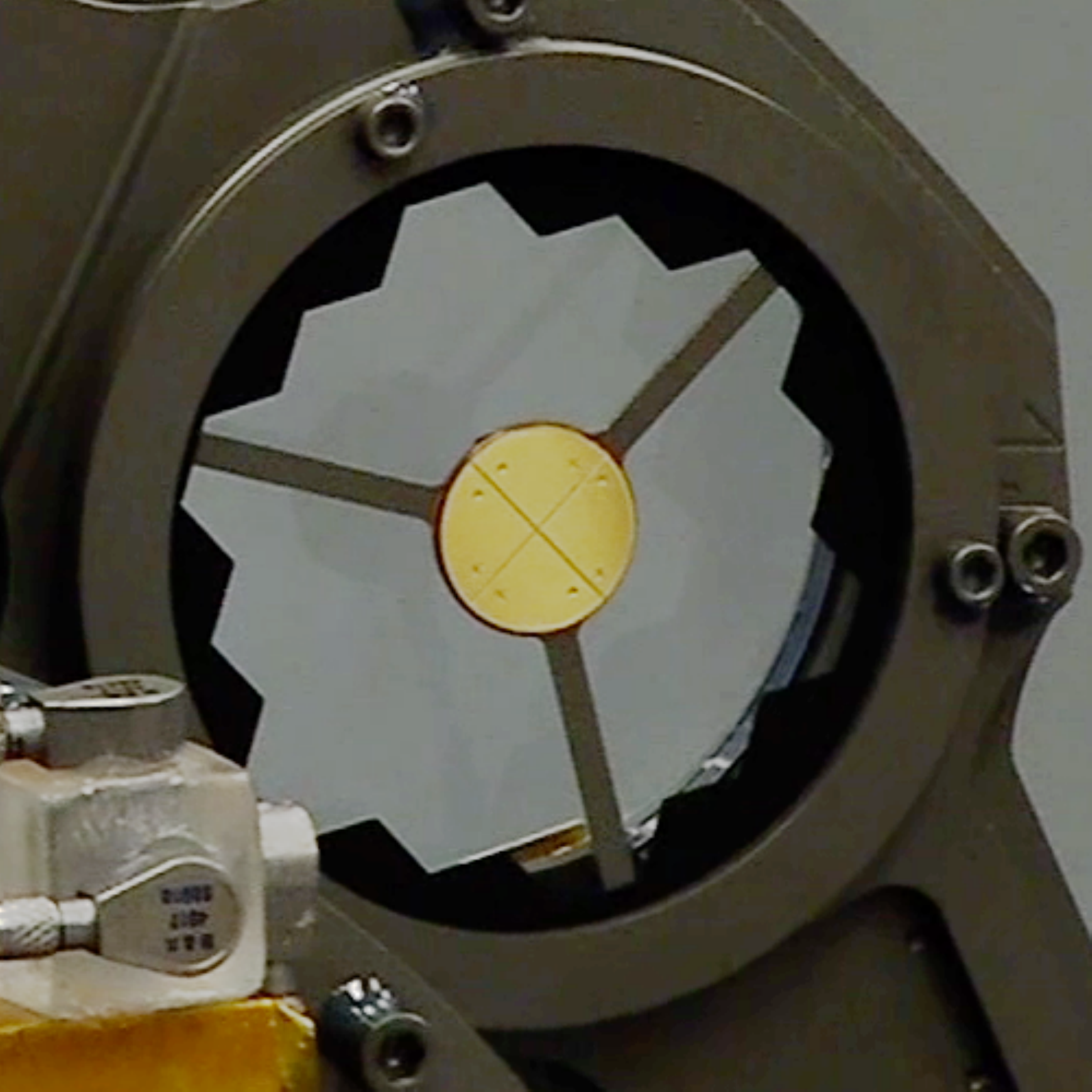}
\caption{
\textbf{Left}: A full size prototype of the NRM flying aboard NIRISS AMI.  The nominal JWST pupil circumscribing circle diameter is  $40~\mathrm{mm}$ in the plane of the mask.  \textbf{Right}: The CLEARP pupil mask enables full pupil images using AMI's filters, target acquisition, and fast readout subarray. The backwards-facing reflective element at the center of CLEARP is an unilluminated pupil alignment mirror which defines the inner edge of the CLEARP pupil.  
CLEARP obscures 5.9\% of the JWST primary mirror's reflective surface.
%
%
Of this, 3.8\% is intercepted by pupil alignment mirror, so incident light is scattered back toward the NIRISS collimator and pick off mirror, while the remainder (2.1\%) is obscured by the structure holding the CLEARP pupil alignment mirror.
Scattering off internal CLEARP edges has not been detected, but is possible.    CLEARP’s outer edge is oversized by $\sim$4\% of the primary mirror's circumscribing circle.}
\label{fig:g7s6clearp}
\end{figure*}

\subsection{Calibration of interferometric data} 
\label{sec:ami_basics}

Atmospheric and instrumental imperfections affect visibilities, which hampers reliable characterization of the astronomical scene. 

For a point source cyclical addition of the fringe phases around a baseline triangle formed by any three subapertures (labelled 1, 2, and 3 below) results in zero  \citep{1958MNRAS.118..276J}:
\begin{eqnarray} 
\label{eq:clopha} 
\Delta\phi_{12} + \Delta\phi_{23} + \Delta\phi_{31} =  \\
(\phi_{1}-\phi_{2}) +  (\phi_{2}-\phi_{3}) + (\phi_{3}-\phi_{1}) &=& 0. \nonumber
\end{eqnarray}
This \textbf{closure phase} identity is used to calibrate out atmospheric and instrumental contributions to phase noise on ground-based optical and near-IR AMI, with or without adaptive optics, as well as data taken with NIRISS AMI, following a radio astronomical precedent applied to optical data.  A similar \textbf{closure amplitude} constraint \citep{1988AJ.....95.1278R} on fringe visibilities measured over baselines from any four subapertures (labelled 1, 2, 3, and $4$) of a point source also exists:
\begin{equation}
\label{eq:cloamp}
    |\cv_{12}|~|\cv_{34}|~/~|\cv_{13}|~|\cv_{24}| = 1.
\end{equation}
Measured closure quantities can be removed from the corresponding combinations of the target to remove atmospheric and instrument contributions to the target's closure quantities.  Unlike fringe visibility amplitudes, fringe (and therefore closure) phases are insensitive to point symmetric structure in the target  (as a consequence of the Hermitian property of the FT, phases are insensitive to symmetric components under inversion).
Closure amplitude calibration has not been explored much on ground-based data, although visibility amplitude calibration is routinely effected by dividing target fringe amplitudes by those of a point source calibrator observed under similar conditions.

The absence of atmospheric instability enables exquisite calibration of interferometric data, as the Hubble Space Telescope (HST) demonstrated \citep{1991ApJ...377L..17F,1992PASP..104..958B}. 
NIRISS AMI's competitive niche compared to ground-based NRMs is engendered by access to better phase calibration as well as unmatched amplitude calibration. 
JWST's  optical aberrations at AMI wavelengths produce $>90$\% Strehl ratio images. Temporal wavefront drift adds only a few nm scale phase creep over time.  Interferometric observables that are inoculated against JWST's phase errors promise contrasts beyond ground-based  capabilities \citep{2022arXiv220705632R}.

\subsection{Kernel phase interferometry}
\label{sec:kernel}

Kernel phase interferometry (KPI) involves analysing images typically formed by the full telescope aperture \citep{2010ApJ...724..464M,2013PASP..125..422M}. It generalizes NRM's Fizeau treatment of data to aperture geometries that are not necessarily sparse. Data acquisition, calibration, and analysis trace the same path as NRM data; {\textit{viz.}} interferometric observables extracted from images are calibrated for instrumental effects using data from a point source reference object acquired under sufficiently similar conditions. The KPI technique has been applied to \emph{HST} data by \citet{Pope2013} to study brown dwarf multiplicity and was first employed for usage from the ground with extreme adaptive optics facilities by \citet{2016MNRAS.455.1647P}. Later, \citet{2019MNRAS.486..639K} and \citet{wallace2020MNRAS.498.1382W} conducted the first KPI surveys for young self-luminous companions at small angular separations which are inaccessible with classical coronagraphy techniques. At the same time, \citet{Laugier2019} developed methods to apply the KPI technique to saturated images and also introduced angular differential kernels \citep{laugier2020A&A...636A..21L} which take advantage of the angular diversity of ground-based pupil tracking mode data to self-calibrate kernel phase observables.

KPI analysis commences with a discrete numerical model of the pupil, by defining a grid of subapertures over the active pupil. Each subaperture pair provides a single (but usually non-unique) baseline in the \uvplane.  Many subaperture pairs can share the same vector baseline, since the pupil is typically redundant. The PSF is written as the intensity of a coherent sum of fringe complex amplitudes from each distinct baseline present in the pupil model. Subaperture pairs in the pupil model are redundant, so closure phase is not a calibratable observable with this pupil.  Instead, when the wavefront error standard deviation is below $\sim 1$ radian, a linear approximation is adopted. 
A singular value decomposition applied to the (approximated) linearized relationship between the set of pupil pistons and the complex visibility $\cv$ retrieves a set of linear combinations of subaperture phases (pistons) that yield self-calibrating observables: kernel phases \citep{2010ApJ...724..464M,2020A&A...636A..72M}.  In common with the familiar closure phases from NRM analysis, these kernel phases are zero for a point source and after calibration they recover robust observables addressing the structure of the science target.  Kernel phases can therefore be used to characterize the astronomical scene after the same fashion as closure phases. As well as kernel phases, there also exists a similar generalization of closure amplitudes (Eq.~\ref{eq:cloamp}) to kernel amplitudes in the small error limit \citep{Pope2016}.

NIRISS KPI observations taken with the AMI template possess the advantage of sub-pixel repeatability of a target and its calibrator (see Section \ref{sec:ops}). A DIRECT observation in this template uses the CLEARP aperture (right panel of Figure \ref{fig:g7s6clearp}). Its vertical strut is superposed on (but wider than) the main JWST  primary's strut. CLEARP's two diagonal struts, on the other hand, do not coincide with the two folding secondary mirror struts. This more complicated CLEARP obscuration must be included in the discrete pupil model. KPI observations can use any of AMI's four allowed filters. CLEARP's  \uvplane\  coverage for each filter is presented in Figure \ref{fig:filtersuv} of the appendix.

NIRISS KPI images taken with the AMI observing template saturate on targets about four magnitudes fainter than AMI's brightest targets, with pre-launch expected contrast ratio limits of $\sim 7.5$ magnitudes \citep{2019A&A...630A.120C}.   NIRISS KPI delivers significantly higher contrast ratios at separations exceeding $\sim350$~mas, outside AMI's OWA.  This overlaps with NIRCam's coronagraphic IWA, so could help tie in NIRISS AMI mode observations with NIRCam's coronagraphic ones. 
 First results from commissioning are reported in \citet{kammerer2022arXiv220800996K} and a detailed description and performance analysis of the NIRISS KPI observing mode can be found in the NIRISS V paper \citep{2022arXiv221017528K}.  The fainter NIRISS KPI targets do not overlap with ground-based ExAO targets, but are, for example,  uniquely suited to an as-yet unexplored search space of binary and multiple ultra-cool dwarf stars.

\subsection{Wavefront sensing with AMI}\label{sec:wavefront_sensing}

The NIRISS AMI PSF contains recoverable wavefront phase and amplitude information. This is calibrated out of science images, but can be used as a wavefront stability monitor.     AMI-like data, taken with sets of co-tilted primary mirror segments  that are in a non-redundant pattern \citep{2009ApJ...700..491M} can be used to drive cophasing JWST. As \citet{2012OExpr..2029457C,2014OExpr..2212924C} show,  any of JWST's science imagers can be used as robust and flexible back-up co-phasing wavefront sensors. For instance, \citet{wong2021JOSAB..38.2465W} demonstrated pupil plane phase retrieval from focal plane images using automatic differentiation methods.  The capture range of such Fizeau wavefront sensing techniques can be extremely large (scaling as the available filter coherence length or wavelength diversity), and it can handle more primary mirror segments than JWST has.  The AMI mode PSF, combined with a full pupil image, breaks the ``twinning'' degeneracy \citep{2012JOSAA..29.2367G} that limits in-focus wavefront sensing.  It is also possible to use pupil asymmetry to break the twinning degeneracy \citep{2013PASP..125..422M,2014MNRAS.440..125P}, although exposure depth required for this on JWST may exceed that of the AMI + full pupil combination.  Combined AMI + full pupil PSFs provide JWST a back-up ``fine phasing'' wavefront sensor \citep{2016OExpr..2415506G,2016ascl.soft10005G}.  We note that every AMI calibrator star observation directly measures the segment-to-segment piston aberrations between the mirror segments used by AMI.
 

\section{AMI Science examples}
\label{sec:scienceexamples}

The scientific rationales behind four  NIRISS Guaranteed Time Observations  AMI and KPI programs are briefly sketched below.

\subsection{Architecture of directly-imaged extrasolar planetary systems}
\label{sec:Binaries}

Among more than 5000 exoplanets detected to date only a few dozen have been directly imaged (e.g. \citealt{2008Sci...322.1348M,lagrange_giant_2010,2013ApJ...779L..26R,macintosh_discovery_2015}).
These are typically young, self luminous giant planets with masses greater than 2\,$M_{\text{J}}$ and separations $\gtrsim 10$\,AU \citep{currie_direct_2022}. In the past 5 years large ground-based ExAO surveys have enabled detailed characterization of the giant planet population at large separations, reaching contrasts down to $\sim$10$^{-5}$ at separations of 0.5 arcsec \citep{nielsen_gemini_2019,vigan_sphere_2021}.
To probe shorter separations aperture masking has been extensively used from the ground. This has led to the detection of several companions and circumstellar disks \citep{hinkley_observational_2011,kraus_lkca_2012,hinkley_discovery_2015,sallum_accreting_2015,sallum_new_2019,2022ApJ...931....3B}.
More recently near-infrared long-baseline interferometry with VLTI/GRAVITY enabled direct detection of planetary-mass companions at separations down to $\sim$3\,AU while delivering astrometry at the level of $100$ $\mu$as \citep{nowak_direct_2020,hinkley_direct_2022}.

Despite covering a somewhat restricted parameter space, direct observations of exoplanets have been extremely valuable to better understand planetary formation, evolution, and atmospheric properties. Photometry and spectroscopy at low, mid, and high resolution have been used to characterize giant planet atmospheres and better understand their formation history \citep{currie_combined_2011,konopacky_detection_2013,chilcote_12.4$upmu$m_2017,wang_detection_2021}.
Moreover, luminosity measurements from imaging can be combined with dynamical mass estimates from radial velocity monitoring \citep{nowak_direct_2020,vandal_dynamical_2020}, host-star astrometry \citep{brandt_first_2021} or non-Keplerian orbital motion in multi-planetary systems \citep{lacour_mass_2021}. This enables constraints on luminosity evolution models, which are not calibrated at young ages, yet are used to infer masses of all directly imaged companions to date \citep{baraffe_evolutionary_2003,marley_luminosity_2007,spiegel_spectral_2012,berardo_evolution_2017}.

Most direct imaging observations to date have been performed at wavelengths below 3\,$\mu$m. In contrast, NIRISS AMI will attain contrasts of 8-9 mag at separations $\lesssim$100-200~mas between 3 and 5\,$\mu$m, where planetary mass companions output most of their light. It will therefore not only enable detection of close-in companions, but also provide precise photometry at these longer wavelengths. Such observations are crucial to better understand the atmosphere physics of these objects and discriminate between theoretical models which, despite predicting similar SEDs at short wavelength, can differ significantly beyond the K band \citep{helling_comparison_2008,allard_bt-settl_2014}.

NIRISS Guaranteed Time Observations (GTO) program 1200 will explore the inner architecture of 3 systems with known planets or debris disks, namely HR 8799 \citep{2008Sci...322.1348M,marois_images_2010,de_rosa_spectroscopic_2016,wahhaj_search_2021,zurlo_orbital_2022}, HD 95086 \citep{2013ApJ...779L..26R,2013ApJ...772L..15R,su_debris_2015,2016ApJ...822L..29R,desgrange_-depth_2022} and HD 115600 \citep{currie_direct_2015,gibbs_vltsphere_2019}. In all these systems, the structure of the disk or the orbital configuration hint towards the presence of yet undetected companion(s) at short separations. AMI therefore represents a unique opportunity to search for putative inner planets while providing precise photometry at 4.8\,$\mu$m.


\subsection{Detecting extrasolar zodiacal light }
\label{sec:zodi}
\noindent

\begin{figure*}[hbt!]
  \centering
  \includegraphics[scale=1]{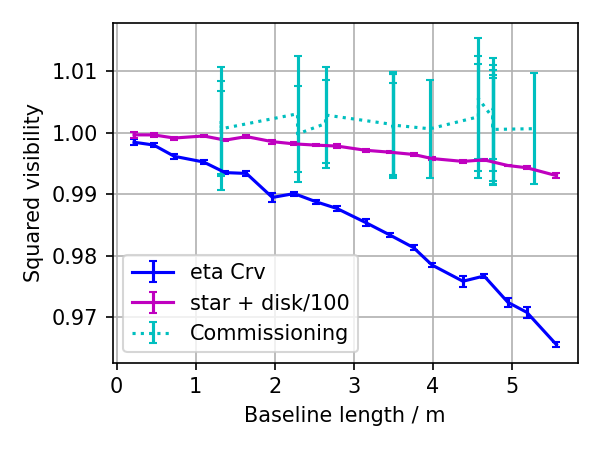}
  \caption{Squared visibilities as a function of maximum baseline length. Simulated AMI observations of the dusty debris system $\eta$ Crv  and a hypothetical exozodiacal system with 100 times lower dust density were analyzed with a precursor to the \texttt{AMICAL} interferometric package.  Commissioning results (utilizing \texttt{ImPlaneIA}) of a calibrator star calibrated against another calibrator star are also shown.  These data indicate $\eta$ Crv's zodiacal light can be detected with AMI.  Improved charge migration calibration could yet render the fainter disk detectable. Such data could constrain the form of the extrazodiacal dust distribution in a routine AMI observation of a single target-calibrator pair.  JWST Program 1242 includes such an observation.}
  \label{fig:zodi}
\end{figure*}

Though likely ubiquitous, dust and debris disks are difficult to detect from the ground despite the Sun's zodiacal disk being $\sim 350$ times brighter than all its planets combined in optical to mid-IR \citep{Defrere2010}.  Only the densest debris systems at inner solar system and habitable zone scales have been detected.  Typical stars are thought to commonly support up to $\sim\,$10\,zodi disks (one `zodi' being the brightness of the Solar zodiacal disk). Such disks hamper or prevent direct imaging of embedded planets \citep{Beichman2006,Kennedy2013}.  15–20\% of all FGK stars support dust systems above the IRAS detection limit of $\sim$100\,zodi, and apparently old ($\sim$\,GYr) systems possess copious dust \citep{Kennedy2013}.  \citet{Beichman2006} report 100--1000\,zodi dusty systems.  CHARA \citep{Ertel2014} and VLTI \citep{Absil2013} have shown that close to stars, hot dust can reprocess up to  1\% of stellar radiation in 30\% and 11\%, respectively, of their surveyed stars.  More recently the LBTI-HOSTS survey \citep{2018AJ....155..194E,2020AJ....159..177E}, reports a $\sim$\,3\,zodi median exozodiacal dust level with an upper boundary of $\sim$\,27\,zodi.  Improved dust and debris disk detection limits, as we predict will arise from the breakthrough in visibility measurement precision (a quantitative simulated example is given in Figure~\ref{fig:zodi}) are critical inputs to future terrestrial planet mission target selection \citep[\eg,][]{stark2014ApJ...795..122S,stark2015ApJ...808..149S,life2022A&A...664A..21Q}.

Planet-induced non-uniformities in a dusty zodiacal disk may be detectable with NIRISS AMI's 100~mas IWA and 4-5\,\nicemicron ~wavelength coverage, providing a view of warm inner-solar-system dust around nearby stars.  Planetary perturbations induce various zodiacal dust signatures (waves, gaps, resonances) on the scale of 100~mas: the size of a habitable zone around a G-type star at 10\,pc. The distribution of this dust is therefore a direct-imaging probe of otherwise unseen planetary systems \citep{Bonsor2018}.

The first planned exozodi observations with AMI will be of $\eta$~Corvi, a 1-2 Gyr old F2V star with a debris disk \citep{Vican2012}. Its spectral energy distribution shows both a hot component constrained by VLTI \& LBTI interferometry to lie between 0.16-2.6\,AU, \citep{Defrere2015,Lebreton2016} and a cold component resolved by ALMA with a radius of 151\,AU \citep{Marino2017}. The persistence of this hot dust despite its short dynamical lifetime has led \citet{Marino2017} and \citet{Marino2018} to suggest cometary material is passed by a `conveyor-belt' of one or more planets from the outer disk to the inner disk. This interpretation is sufficiently dynamically favoured that it may explain hot dust populations in many other systems, but remains controversial \citep{Pearce2022}. AMI interferometry of $\eta$~Crv in GO~1242 (simulated in Figure~\ref{fig:zodi}) will probe the spatial scales in between the under-resolved LBTI and over-resolved VLTI to make the first map of the hot inner disk.

\subsection{Feedback in active galactic nuclei }
\label{sec:agn}
\noindent
AGN are thought to be key ingredients in driving structure formation and supermassive black hole mergers, but there are many unknowns when it comes to fuelling AGN.  By clearly resolving structures in the outskirts of the archetypal Type 2 Seyfert galaxy NGC 1068 (GTO Program 1260), AMI can help answer some of these important questions. Very recently, ground-based interferometric observations have provided new insights on the structure of the dust distribution around the core of NGC 1068.  In the L-band, with MATISSE, \citet{2022Natur.602..403G} report a two component structure, an optically thick ring obscuring the central engine at parsec scales and a less optically thick disk extending to at least 10 pc. In K band, \citet{2020A&A...634A...1G} reported the presence of an inner thin ring with a radius of only 0.24\,pc, consistent with the presence of dust at a sublimation temperature of T = 1500 K using the VLTI-GRAVITY interferometer. The component found with the GRAVITY data could be interpreted as the inner most component reported by \citet{2022Natur.602..403G}. However, the best-fitted models to the GRAVITY data favour the presence of a flat ring-like structure, instead of tori. Very recently,  \citet{2022ApJ...926..192V} reported models of the NGC~1068 SED in N and Q bands, favouring the two components scenario reported with the MATISSE data. The SED fitting supports a very complex dust morphology formed mostly by a silicate+carbon composite with large (1~$\mu$m) dust-grain sizes. Further analysis of interferometric data might help to better understand the complexity of the morphology of the core of NGC~1068, in this regard complementary AMI data with the JWST will be of particular importance.
AMI in the F480M bandpass targets a region 5~--35~pc from the nucleus \citep{2014ApJ...783...73F}, where inflowing or outflowing warm dust may occur.  Large-scale streamers fuelling the parsec-scale ring would suggest that what we think of as parsec-scale AGN can be fuelled for a long time by ongoing instabilities and turbulence in much more distant (many pc scale) reservoirs.  Conversely, the absence of such connections might suggest that AGN are a one-off episode on parsec scales, with implications for how much feedback could actually be sustained by such systems. The presence or absence of heated warm gas in the polar directions from NGC 1068 will put strong constraints on the strength of outflows and feedback expected from the poles. AMI also ties detector astrometry to the Gaia coordinate system, which would place ground-based L-band interferometry of NGC~1068 on an absolute astrometric frame.

\subsection{Transition disk science}
\label{sec:TD}

Millimeter interferometers reveal detailed structure within the circumstellar disks surrounding young, forming stars \citep{Andrews_2018, Long_2018}. A few of these protoplanetary disk systems, termed transition disks, display inner gaps and holes suggesting that they are on the verge of losing their natal disk material \citep{Espaillat_2014,Francis2020}, potentially influenced by planet driven dynamical clearing. Transition disks are thus excellent hunting grounds for finding and characterizing forming planets. The extreme youth of these systems ($\sim$Myrs) and the paucity of dust emission within their gaps make them ideal targets to search for (proto)planets. Infrared observations are preferred as young (proto)planets will be bright in the thermal infrared due to both their inherent thermal emission and the addition of a planetary accretion disk and a possible accretion shock \citep[e.g.][]{Sallum2015}. The observed inner hole sizes for Transition Disks range from 1 -- 100\,au for the nearest protostars 100 -- 150\,pc from the Sun. NIRISS AMI is extremely well suited for finding and characterizing these forming (proto)planets and GTO Program 1242 will study three nearby systems: PDS\,70, HD\,100546, and HD\,135344B. 

The sparse nature of AMI  observables, compared with the possibly complex observing scene, consisting of a central star, potential (proto)planet, and extended disk emission complicate the analysis and can make it difficult to distinguish between planet and disk emission \citep{Cieza_2012,Kraus_2013}. To achieve success, it will be important to model the extended disk emission while simultaneously fitting for planets \citep{2022ApJ...931....3B}. 
\begin{figure*}[ht]
  \centering
  \includegraphics[scale=0.500]{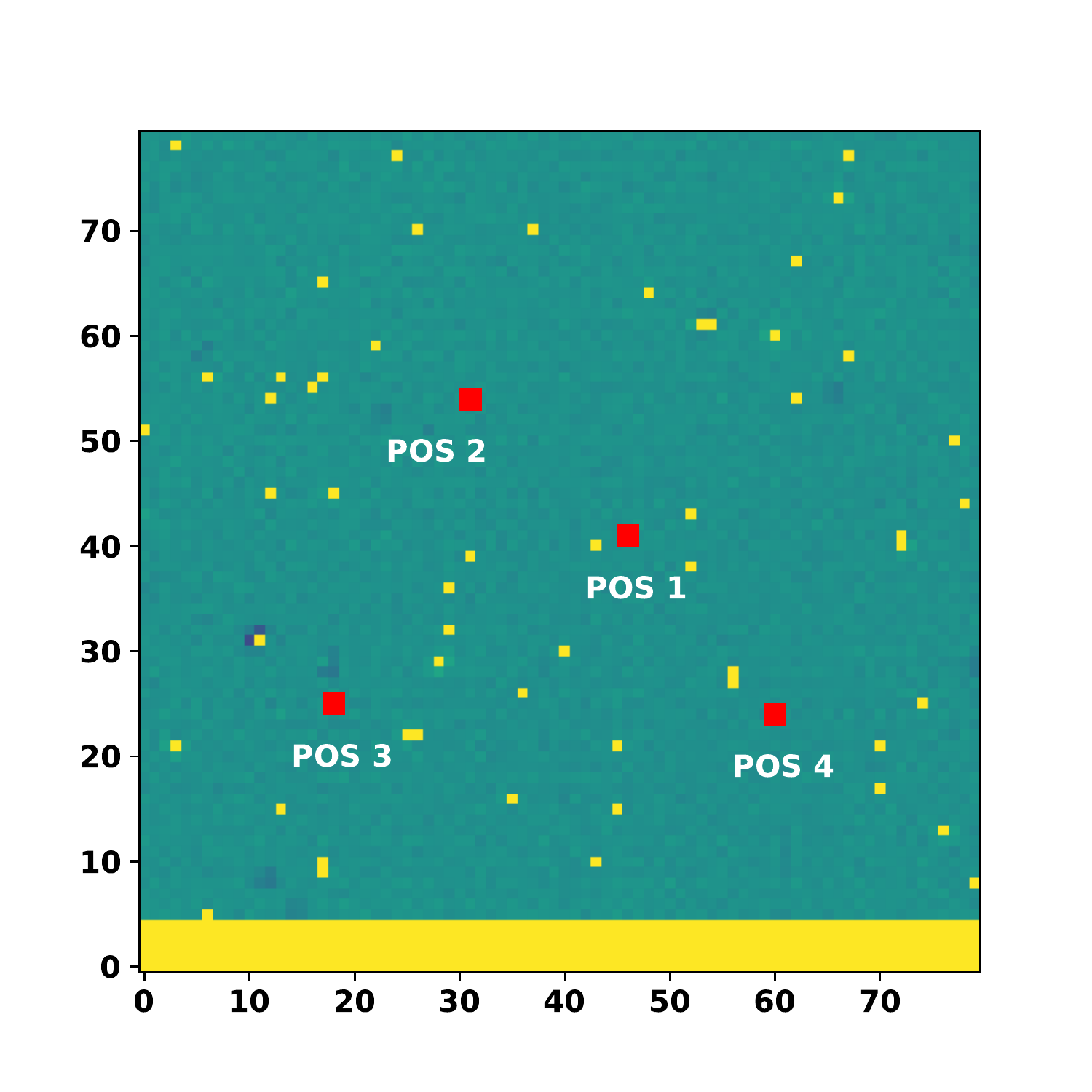}
  \caption{Four target locations (in red) on the AMI SUB80 subarray flat field,  with known persistent bad pixels and 4 rows of light-insensitive reference pixels (in yellow).  Most programs should only use POS 1, and avoid dithering. NIRISS KPI observations use the same POS1 location as NRM observations.}
  \label{fig:F480Flat}
\end{figure*}
%

\section{Operations}
\label{sec:ops}

The STScI Astronomer's Proposal Tool (APT) implements an AMI observing template to plan NIRISS  AMI and KPI observations.  Such observations are  more easily parameterized by the number of photons collected, rather than a traditional signal-to-noise ratio of the peak pixel used in the JWST exposure time calculator \citep{2013MNRAS.433.1718I}.  

\subsection{Subarray and dithers}
\begin{figure*}[ht!]
  \centering
  \includegraphics[scale=0.500]{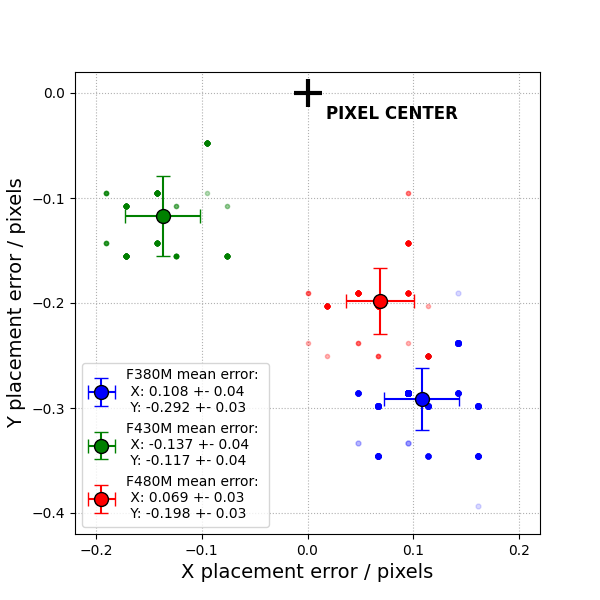}
  \caption{AMI target acquisition accuracy.
     Every AMI mode observation commences with target placement at the center of SUB80's POS1 pixel, ideally, as a mitigation against flat fielding error.  A single target acquisition is used for the three medium band AMI filters' observations during commissioning.  Here placement errors for all NRM commissioning target acquisitions are plotted using semitransparent points.  These $\sim$0.3 pixels (20 mas) errors are dominated by filter-dependent offsets. The error can be reduced to $\sim$2 mas, the scatter around filter-specific offsets if each filter is placed in its own observation (with the  associated temporal overhead for target acquisition).  This would enable filter-specific offsets at the cost of efficiency.  If the three filters are used in a single observation, commissioning results suggest an offset of $\sim$0.2 pixels in the Y direction would improve target centering. A precomputed grid of oversampled PSFs was used to determine the image locations \citep{2000PASP..112.1360A}.
  }
  \label{fig:TAaccuracy}
\end{figure*}

The AMI template is optimized to reduce the mode's sensitivity to flat fielding errors and possible sensitivity variations within a pixel, so compact targets are placed at the center of a predefined pixel on the detector (POS 1). Using the POS 1 position for most AMI observations will help calibrate pixel response in relevant pixels. Four rows of light-insensitive edge pixels help calibrate out temporal gain drifts, making the field of view 5.2 x 5 arcseconds. This SUB80 subarray is continuously read out every 0.07544\,s (TFRAME), which enables AMI to observe bright extreme adaptive optics targets.
While dithering is discouraged, it is an option.
The AMI template offers four different fixed primary dither positions (see Figure \ref{fig:F480Flat}) as well as up to 25 small sub-pixel sub-dithers around the primary positions\footnote{\href{https://jwst-docs.stsci.edu/jwst-near-infrared-imager-and-slitless-spectrograph/niriss-operations/niriss-dithers/niriss-ami-dithers}{NIRISS AMI dithers}}.

An offset can also be applied to the entire observation, which makes it possible to put a target anywhere within SUB80.  The offset can be chosen to place an acquired target at the center of the POS1 pixel in a particular filter.  The current AMI observing template offers one offset for an observation.   In order to center two or more filters' NRM images to $\sim$2~mas  observations, each filter will require a separate observation with its own filter-specific offset  
\citep{1982JOSA...72.1199T,1995AJ....110..430S,JWST-STScI-0003965H}.

\subsection{Target acquisition}
A target is placed at a specified pixel, to sub-pixel accuracy, using AMI's target acquisition (TA)\footnote{
    \href{https://jwst-docs.stsci.edu/jwst-near-infrared-imager-and-slitless-spectrograph/niriss-operations/niriss-target-acquisition}{NIRISS target acquisition}} sequence and the F480M filter \citep{JWST-STScI-0003963H,JWST-STScI-0003965H}.
After telescope slew, the target is placed in the 64x64 SUBTAAMI subarray adjacent to SUB80. On-board software then determines the target location using a few exposures, and a small angle maneuver moves the target to its desired location on SUB80. This maneuver places the NRM+F480M PSF to within 0.25 pixels from the pixel center of position POS1 (Figure \ref{fig:TAaccuracy}). Keeping compact sources away from pixel edges should reduce susceptibility to pointing jitter, intrapixel sensitivity, and charge migration effects. The NRM is used  for TA
when targets are brighter than 9.3 mag (AMIBRIGHT), otherwise CLEARP is used (AMIFAINT), even if the NRM is used for observing. TA may fail if a target is fainter than $M=14.5$.  For kernel phase observations, after TA,  an initial NRM exposure of the science target is currently required 
before CLEARP can be inserted for the remaining observations (specified in the \textit{Direct Imaging Parameters} section of the APT observation template). The pupil wheel is not moved during an observation because its exact position has an uncertainty of 0.157\degr. Should TA fail, the observation will still be executed under the assumption the target was at the last location in the TA subarray where centroiding was attempted.  A small angle maneuver then moves this assumed target location to the first science position in the SUB80 array.  We expect the target will still be placed within the detector, and the science observation will be carried out.

JWST's allowed roll angle\footnote{\href{https://jwst-docs.stsci.edu/jppom/special-requirements/aperture-position-angle-special-requirements}{Position Angle special requirements}} is highly constrained. In consequence, the sun angle determines the general pointing of the telescope. Filling in gaps in the NRM's \uvplane\ coverage is best done, when possible, by observing a target on different dates, rather than by rotating the telescope.

\subsection{PSF Reference Star as a Calibrator}

A reference image of an unresolved single star is currently recommended for programs requiring contrasts exceeding $\sim$10:1. This reference star should be single and of similar brightness and spectral type as the science target.
A target and its calibrator are best observed with minimal delay between them, and they should be close to each other in order to minimize telescope and instrument thermal changes between two observations.
Reference star selection often involves using existing tools and catalogs. These considerations are discussed in more detail in the AMI documentation\footnote{\href{https://jwst-docs.stsci.edu/jwst-near-infrared-imager-and-slitless-spectrograph/niriss-observing-strategies/niriss-ami-recommended-strategies}{NIRISS AMI recommended strategies}}. In the future, a library of calibrator stars' PSFs might enable higher contrast with fewer calibration observations. 
        
\section{Data analysis procedure}
\label{sec:analysis}

\subsection{AMI mode data and the JWST Pipeline} 
\label{sec:pipeline}

JWST's Science Calibration Pipeline processes raw data into calibrated data products. It typically consists of 3 main stages.  Each stage has default settings, though multiple customizable steps can be run individually.

The first stage---\texttt{calwebb\_detector1}---applies detector-level corrections to all data before performing ramp fitting.  To avoid complicating interferometric calibration interpixel capacitance correction (\texttt{ipc}) is, as a default, not run on AMI mode data.  Next, \texttt{calwebb\_detector1} outputs corrected countrate image (\texttt{rate}) files and their per-integration counterpart (\texttt{rateints}) files.

The second stage treats imaging and spectroscopy separately. \texttt{calwebb\_image2} is used on AMI mode data. This stage performs additional physical (e.g., flux calibration) and instrument (e.g., flat-fielding) corrections and produces calibrated (\texttt{cal}) files and per-integration (\texttt{calints}) images. AMI mode processing skips photometric calibration (\texttt{photom}) and resampling (\texttt{resample}) by default.

Stage 3 NIRISS AMI and KPI processing is not currently done in the pipeline, but in free-standing software described in Section \ref{sec:extract} and Section \ref{sec:kernel_analysis} below.

\subsection{Bad pixels in AMI and KPI data }
\label{sec:bad_pixels}

Pipeline stages 1 and 2 identify some bad pixels.  A Fourier power minimization algorithm to correct bad pixel values \citep{2013MNRAS.433.1718I} is used (instead of convolution, median filtering, or interpolation) on AMI mode data.  This method has been used successfully on Nyquist or finer image sampling of full pupil data, and we are here applying it to NRM data. We note that AMI is Nyquist sampled at 3.8~\nicemicron~(F380M) and beyond but undersampled in F277W for which the discussed algorithm might yield reduced performance.


The basic principle of the \citet{2013MNRAS.433.1718I} method is that isolated saturated or bad pixels generate spurious signal that in theory lies outside the support of the visibility (which is also an optical system's OTF support). Removing such unphysical power from a numerical visibility array can be achieved by adapting the bad pixels' values which is a linear problem, though for reasons of robustness it is done iteratively \citep{2019MNRAS.486..639K}. Saturated pixels in the core of an image can also be corrected this way.  Bad pixels not flagged by the JWST pipeline are identified and corrected in the course of the iterations. A more detailed mathematical description of this minimization problem can be found in \citet{2013MNRAS.433.1718I} and \citet{2019MNRAS.486..639K}. The minimization problem is solved using a cropped image centered on the target, so correction is restricted by the smallest pixel row or column distance to the edge of the data frame.

\subsection{Interferometric Observable Extraction}
\label{sec:extract}

After bad pixels are corrected, the cropped individual integration image arrays are input to software that extracts raw fringe phases and amplitudes and their derived closure quantities, and, in AMI case, segment piston phases as well.  
There are two common ways to extract these Fourier quantities from AMI images. The first one fits a model PSF (or interferogram) to the images in order to derive fringe phases and amplitudes for all baselines of the NRM. The second one does not require a model interferogram and instead numerically Fourier transforms the images themselves to obtain complex visibilities for all baselines.
AMI's Nyquist or coarser pixel pitch encouraged the decision to develop an analytical fringe model \citep{2011A&A...532A..72L,2005ApOpt..44.1360S,2015ApJ...798...68G,2018ascl.soft08004G}, 
which is insensitive to image cropping.  In contrast,
numerical fringe extraction effectively convolves true visibilities by the data windows' FT and can suffer from aliasing effects if the data is undersampled (such as in F277W).

The third stage of JWST AMI data processing does not use the JWST pipeline---it uses a JWST AMI-specific branch\footnote{\url{https://github.com/anand0xff/ImPlaneIA}} of  \texttt{ImPlaneIA} \citep{2018ascl.soft08004G} instead.
Steps in this extraction are:
\begin{enumerate}
    \item Crop each frame of JWST stage 2 pipeline output (\texttt{calints} file) around the peak of the image.
    \item Create the 44 component fringe model (Equation \ref{ae}) using the as-designed pupil geometry.
    \item Measure image centering and rotation, then determine the best-fit model coefficients.
    \item Output raw observables derived from model coefficients (using Equations \ref{ad} and \ref{ad2})  in OIFITS v2 format \citep{2017A&A...597A...8D}, the image plane model and residuals as 2d FITS images, and observables in text format. 
    \item Calibrate target OIFITS file(s) with calibrator star OIFITS file(s) to create calibrated OIFITS output.
\end{enumerate}
ImPlaneIA saves a single set of observables and their standard deviations as well individual integrations' observables.  A frame selection utility to identify and reject individual frames is being developed.

Two packages, \texttt{AMICAL} and \texttt{SAMPip}, offer alternatives to \texttt{ImPlaneIA}.  All three python packages have run on NIRISS AMI commissioning\footnote{JWST Program 1093 (PI: D. Thatte)} as well as JWST Early Release Science NIRISS AMI\footnote{JWST Program 1349 (PI: R. Lau)} data.
They all read in stage 2 JWST pipeline output and write observables in OIFITS v2 format.

\texttt{AMICAL}\footnote{\url{https://github.com/SAIL-Labs/AMICAL}} \citep{soulain_james_2020} uses a numerical Fourier transform to extract of interferometric observables. It computes the expected position of the peaks in the visibility amplitude $|\cv(u,v)|$  for each pair of apertures, assuming a mask design, observing wavelength, filter bandpass, and detector pixel scale. Various ways of accounting for the difference between the mask design and its implementation are used in order to increase the accuracy of this numerical transform approach, which makes it less sensitive to uncalibrated pupil distortion, for example. 
\texttt{AMICAL} provides end-to-end software  that  (1): applies its own background subtraction, cropping, cleaning of residual bad pixels and cosmic rays, and data windowing to JWST stage 2 pipeline output, (2): performs $\sigma$-clipping and frame selection, (3): determines the uncertainties of the observables, and (4): calibrates the observables with a designated calibrator star's observables.  \texttt{AMICAL}  extracts observables much faster than \texttt{ImPlaneIA} does, since it does not perform a least squares fit for a fringe model.  Image plane data windowing that is too narrow can affect the numerical Fourier transform. 
\texttt{AMICAL} encapsulates two decades of NRM algorithm development dedicated to ground-based NRM. \texttt{AMICAL} is customized for NRM data from JWST NIRISS, ESO/SPHERE, ESO/VISIR, and Subaru/VAMPIRES. A recent demonstration of its performance on ground-based data can be found in \citep{2022ApJ...931....3B}.

\texttt{SAMPip}\footnote{\url{https://cosmosz5.github.io/CASSINI/about/}}, like \texttt{ImPlaneIA},  uses a fringe fitting algorithm. \texttt{SAMPip} implements a \textit{sinc} function to account for finite bandpass smearing of the interferometric fringes, which enables it to run faster than \texttt{ImPlaneIA}.  It too incorporates frame selection based on the stability of the total counts in the stack of integrations presented in \texttt{calints} FITS files from JWST's \texttt{calwebb\_image2} pipeline. 

\subsection{Kernel Phase Extraction}
\label{sec:kernel_analysis}

\begin{figure*}[ht!]
  \centering
  \includegraphics[width=\textwidth]{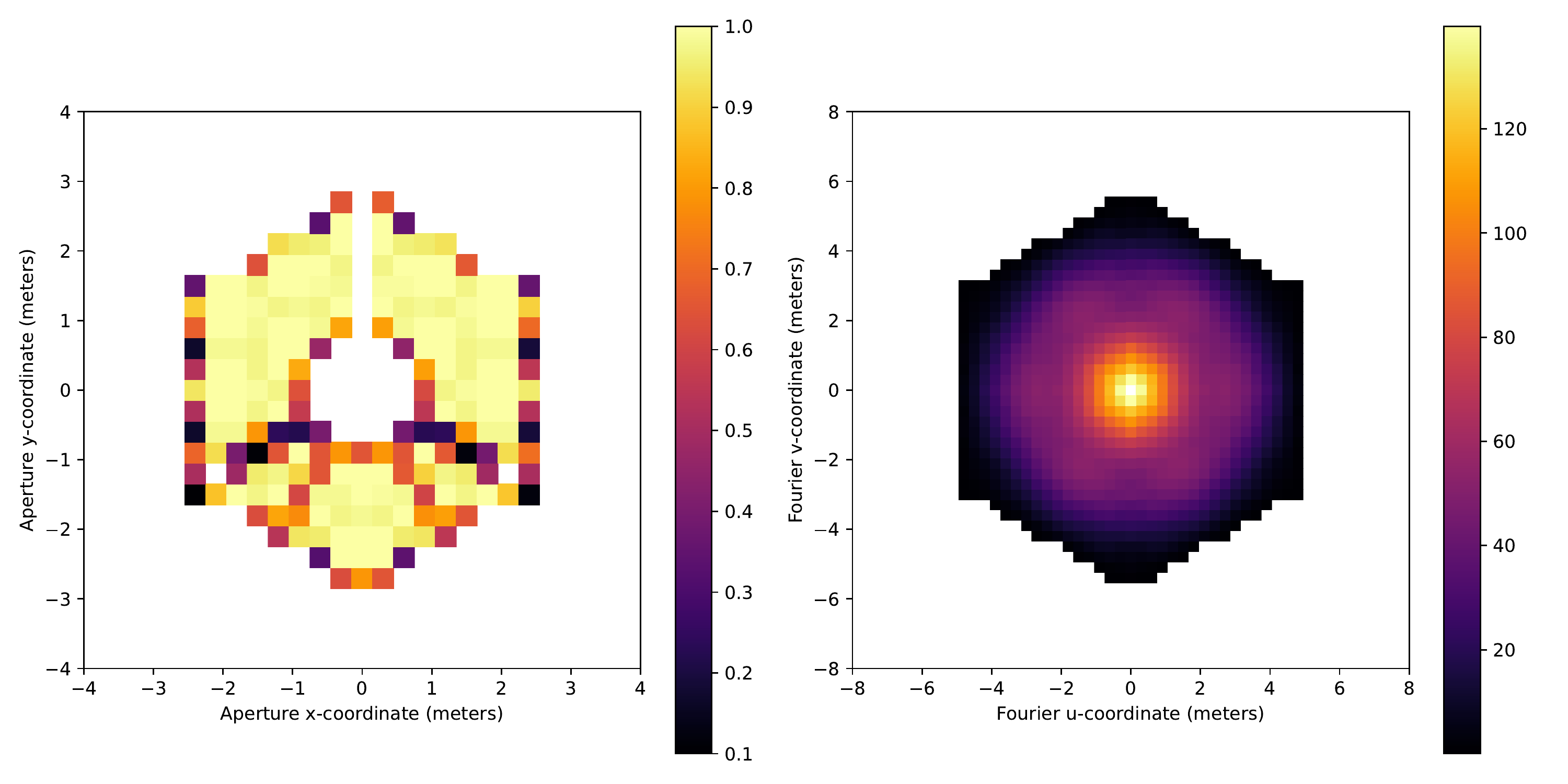}
  \caption{Left: An example of a NIRISS CLEARP pupil model.  Finer scale models exist (\eg,~Kammerer et al. 2022, accepted).  The ``grey'' pupil model contains partially transmissive subapertures. Right: Fourier plane coverage of the pupil model on the left. The color scale encodes the redundancy (\ie,  multiplicity) of each baseline.}
  \label{fig:pupil_model}
\end{figure*}

Kernel phase extraction is performed in the \uvplane,  following \cite{2010ApJ...724..464M}, as implemented in the \texttt{xara}\footnote{\url{https://github.com/fmartinache/xara}} package \citep{2010ApJ...724..464M,2013PASP..125..422M,2020A&A...636A..72M}. For JWST KPI a kernel phase-specific stage 3 pipeline, \texttt{Kpi3Pipeline}\footnote{\url{https://github.com/kammerje/jwst-kpi}}, was developed based on \texttt{xara}. Its interface mimics that of STScI's \texttt{jwst} stage 3 pipelines for coronagraphy. \texttt{Kpi3Pipeline} accepts NIRISS, and also NIRCam, full pupil images (stage 2 pipeline outputs). It consists of three additional steps, following the HST/NICMOS kernel phase analysis strategy \citep{2010ApJ...724..464M,Pope2013,2020A&A...636A..72M} as well as several ground-based kernel phase programs \citep[\eg,][]{2014MNRAS.440..125P,2016MNRAS.455.1647P,2019MNRAS.486..639K,Laugier2019}:
\begin{enumerate}
    \item Re-center individual frames to subpixel accuracy by minimizing the Fourier phase slope.
    \item Apodize the frames with a super-Gaussian function to reduce numerical FT edge artifacts.
    \item Extract kernel phases using a linear discrete FT at the subaperture positions defined by the pupil model.
\end{enumerate}
Unlike the stage 3 AMI pipeline, the \texttt{Kpi3Pipeline} is not formally integrated into the JWST data reduction pipeline and thus has to be run manually by the user on the stage 2-reduced images.

We note that the re-centering is done in the Fourier plane once the complex visibilities have been extracted instead of in the image plane itself so to avoid introducing pixel interpolation errors. The super-Gaussian function's scale (or ``radius'') is governed by the frame size. In practice, the smallest cropped, centered frame extracted from the SUB80 array has a 20~pixels super-Gaussian scale. The pipeline includes a default discrete pupil model. The undersized pupil model used on commissioning data is 85\% of the NIRISS CLEARP pupil. This excludes the longest baselines, which carry the least signal and may hence be noisy. Figure~\ref{fig:pupil_model} shows this pupil model together with its \uvplane\ coverage. Subapertures are on a pitch of 0.3~m.

\texttt{Kpi3Pipeline} output is in kernel phase FITS files (KPFITS) format. 
This pipeline, its use with NIRISS observations, and the KPFITS format are described in NIRISS paper V (Kammerer et al. 2022, accepted). 

\subsection{Modelling the astronomical scene}
\label{sec:modelling}

With complex visibilities and closure or kernel phases in hand,  both AMI and KPI follow the same procedure. A parametric model is first used to predict the visibility signal induced by an assumed scene geometry.
The model's visibility is then projected into the space of closure or kernel phases, and the model's observables are compared to those of the (typically calibrated) data. 

Several packages can fit multiple point sources to data. The \texttt{fouriever} toolkit (Kammerer et al. 2022, accepted) implements a binary-fitting algorithm like that in CANDID \citep{2015A&A...579A..68G}\footnote{\url{https://github.com/amerand/CANDID}}, but can take into account correlation between observables \citep{ 2019MNRAS.486..639K, 2020A&A...644A.110K}. \texttt{fouriever} can compute detection limits and estimate the best fit parameters 
and parameter posterior distribution with Markov Chain Monte-Carlo (MCMC).  To estimate binary parameters, least-squares minimization  over a range of binary contrasts, on a customizable grid of relative source locations is performed.  This produces a $\chi^{2}$ map of local minima. Then the log likelihood function is maximized to find the best-fit binary model, after which its statistical significance is evaluated.  Finally, detection limits are computed by the injection method: after analytically removing the best-fit binary model, artificial companions are injected and their parameters retrieved in order to provide a $3\sigma$ companion detection.  A second contrast limit map is also created using the method in \cite{2011A&A...535A..68A} by comparing a single star model to a binary model and computing the probability that the binary model is consistent with the data. The injection method is considered less sensitive to poorly constrained errors \citep{2015A&A...579A..68G}.  \texttt{fouriever} can be used for either AMI or KPI (as well as long-baseline interferometry data), and was tested on NIRISS commissioning data  (see Section \ref{sec:perf}).

While MCMC can be used to sample the posterior distribution of model parameters, it does not provide a simple metric to compare different likely models. \citet{2022ApJ...931....3B} used nested sampling to enable model comparison with Bayesian evidence \citep{skilling_nested_2006}.


\section{Preparing AMI proposals}
\label{sec:prep4prop}

Preparing AMI proposals to observe binary point source targets is simpler than doing so for extended objects, since modeling binaries involves very few parameters.  Effective proposals for more complicated targets require an appreciation of technical image reconstruction goals that are needed to answer a proposal's driving science questions.  In this section we describe AMI mode data simulators, binary point source exposure time calculation, and an example of how the first extended object AMI proposal was developed using optical bench AMI PSFs that predated today's realistic numerical JWST PSF simulators. 

\subsection{Numerical data simulators}

The STScI-supported Python package Multi-Instrument Ramp Generator \citep[\mirage,][]{2022ascl.soft03008H} can simulate  AMI mode data described in a JWST Astronomer’s Proposal Tool (APT) file.  Combining input source catalogues and YAML configuration files, it creates MAST-format data files.  \mirage~  utilities can produce needed YAML input files for each exposure in the APT file. Current JWST Calibration Reference Data System (CRDS) reference files specify characteristics of detector noise such as read noise, flat field, and dark current.

The alternative method, \texttt{ami\_sim}\footnote{\url{https://github.com/anand0xff/ami_sim}}, is not coupled to APT files. Its simpler architecture mirrors NIRCam image simulations \citep{2003SPIE.4850..388S,2004SPIE.5487..909S}. It convolves user-supplied PSF and sky scene files, adds estimated instrument and pointing noise, and outputs a cube of 2-dimensional integrations. 
Noise characteristics can be set with arguments to \texttt{ami\_sim}, and the code can be modified by the user.  \texttt{ami\_sim}'s input PSF FITS file can be simulated  
analytically \citep[\eg,~\texttt{ImPlaneIA},][]{2018ascl.soft08004G}
or numerically \citep[\eg,~WebbPSF,][]{2012SPIE.8442E..3DP}.
The AMI NRM in WebbPSF's supporting data enables its numerical unity-peak PSF to lie within $ \sim 10^{-5}$ of the analytical value. Either approach should suffice for AMI's $10^{-4}$ contrast detection (at 5-$\sigma)$ requirement \citep{TR-JWST-STScI-007417}.

\mirage~ outputs MAST-format uncalibrated data or linearized ramps.  The latter have had detector-level effects removed. \texttt{ami\_sim} only includes noise expected to remain after the JWST Detector1 pipeline is run. \mirage~  incorporates optical distortion effects seamlessly should they be present in the PSFs from WebbPSF's ``gridded PSF library''. \texttt{ami\_sim} output does not need further level 1 or 2 pipeline processing, but additional keywords from primary and science headers need to be added before extracting observables. \texttt{ImPlaneIA} can graft the required header information from a sample JWST MAST-format file onto {ami\_sim} output; some keyword values (\eg, filter name and telescope roll angle) may need to be adjusted to describe the data appropriately prior to extracting observables.
\texttt{ami\_sim} is suited to an agile exploration of possible target structures or contrasts, noise characteristics, or a range of science cases, whereas \mirage~  is suited to more focused proposals, and for developing bespoke data analysis methods.

\subsection{Binary point source exposure times}


Binary point source exposure times depend on the desired contrast and the fundamental limit set by the closure phase measurement accuracy:  
\begin{equation}
    \sigma_{\text{CP}} = \frac{N_{\text{h}}}{N_{\text{p}}V}\sqrt{1.5(N_{\text{p}}+N_{\text{b}}+n_{\text{p}}\sigma_{\text{ro}})} ~,
\end{equation}
where $N_{\text{h}}$ is the number of holes in the NRM, $N_{\text{p}}$ is the number of collected photons in the interferogram, $V$ is the fringe visibility, $N_{\text{b}}$ is the number of collected background photons, $n_{\text{p}}$ is the number of pixels, and $\sigma_{\text{ro}}$ describes the detector readout noise \citep{2013MNRAS.433.1718I}. Assuming that both background and readout noise are negligible and that a point source is observed (\ie, $V = 1$),  the required number of collected photons is a function of the desired contrast performance
\begin{equation}
\label{eqn:contrast_limit}
    N_{\text{p}} = 1.5\cdot7^2/\sigma_{\text{CP}}^2 \approx 100/\text{contrast}^2,
\end{equation}
where $1.5\cdot7^2$ is conservatively rounded upwards to 100. $\sigma_{\text{CP}}$ (in radians) is equivalent to the achievable 1-$\sigma$ contrast in the high-contrast regime. Published systematic noise floor explorations \citep{2015ApJ...798...68G} were  compared with GPI aperture masking observations \citep{2016A&A...590A..90L,2019AJ....157..249G} and on AMI data simulations.  Unknown or uncontrollable  systematics of NIRISS AMI have been found to reduce achievable contrasts by $\sim$0.5--1 magnitude worse than the theoretical limit.  Improved data calibration may bring AMI accuracy closer to theoretical limits in future.

\subsection{Extended object exposure times}

JWST Program 1260 on the AGN NGC 1068 discussed in Section \ref{sec:agn} was the first extended object AMI proposal to be developed. The process utilized testbed AMI PSFs to simulate observing a relevant range of target morphologies, and an investigation of fringe calibration approaches.

Input PSFs were obtained from the Stony Brook NRM testbed \citep{2011StonyBrookThesisP}, which generated data using an NRM with AMI's pupil geometry. Raw point source closure phase and closure amplitude standard deviations of these data were $0.14\degr$ and $0.0025$ respectively,  which nominally support $\sim$500:1 binary point source detection without calibration. These PSFs' stability provided some confidence for their use as a proxy for space-based observations. One testbed PSF was used to simulate observing NGC 1068, and a different PSF for the simulated AMI calibrator observation \citep{2014ApJ...783...73F}.  Currently numerical PSF simulations validated against JWST data are more convenient for data simulation.

A 7-hole NRM nominally provides 21 point \uvplane~ coverage.  However, testbed stability enabled hundreds of calibrated complex visibilities to be recovered from the visibility array.  This array covers the \uvplane~ more fully than a pinhole representation of the mask, as seen in the rightmost panel of Figure \ref{fig:pup_psf_mtf}---strict non-redundancy was sacrificed to enable expanded coverage.
The best image reconstructions in \citet{2014ApJ...783...73F} used calibrated fringe (\ie~ Fourier) rather than closure phases in CLEAN \citep{2011ascl.soft06007S,1974A&AS...15..417H}.  Subtracting de-sloped calibrator PSF's Fourier phases from de-sloped science target phases, and dividing target fringe amplitudes by calibrator fringe amplitudes produced complex visibilities over much of the \uvplane  (de-sloping a numerical complex visibility array is equivalent to centering on the image's intensity centroid).

Image reconstruction performed on numerically simulated observations of NGC 1068 models provided the required exposure depth ($\sim 10^7$ photons) to distinguish between interesting differences in the input models.  Given uncertainties in JWST performance, this early proposal aims to collect $10^8$ photons.  To fill in subaperture geometry-induced gaps in the \uvplane~coverage two telescope orientations differing by more than 15 degrees are planned.


\section{Performance during commissioning}
\label{sec:perf}

\begin{figure*}[hbt!]
  \centering
  \includegraphics[scale=0.550]{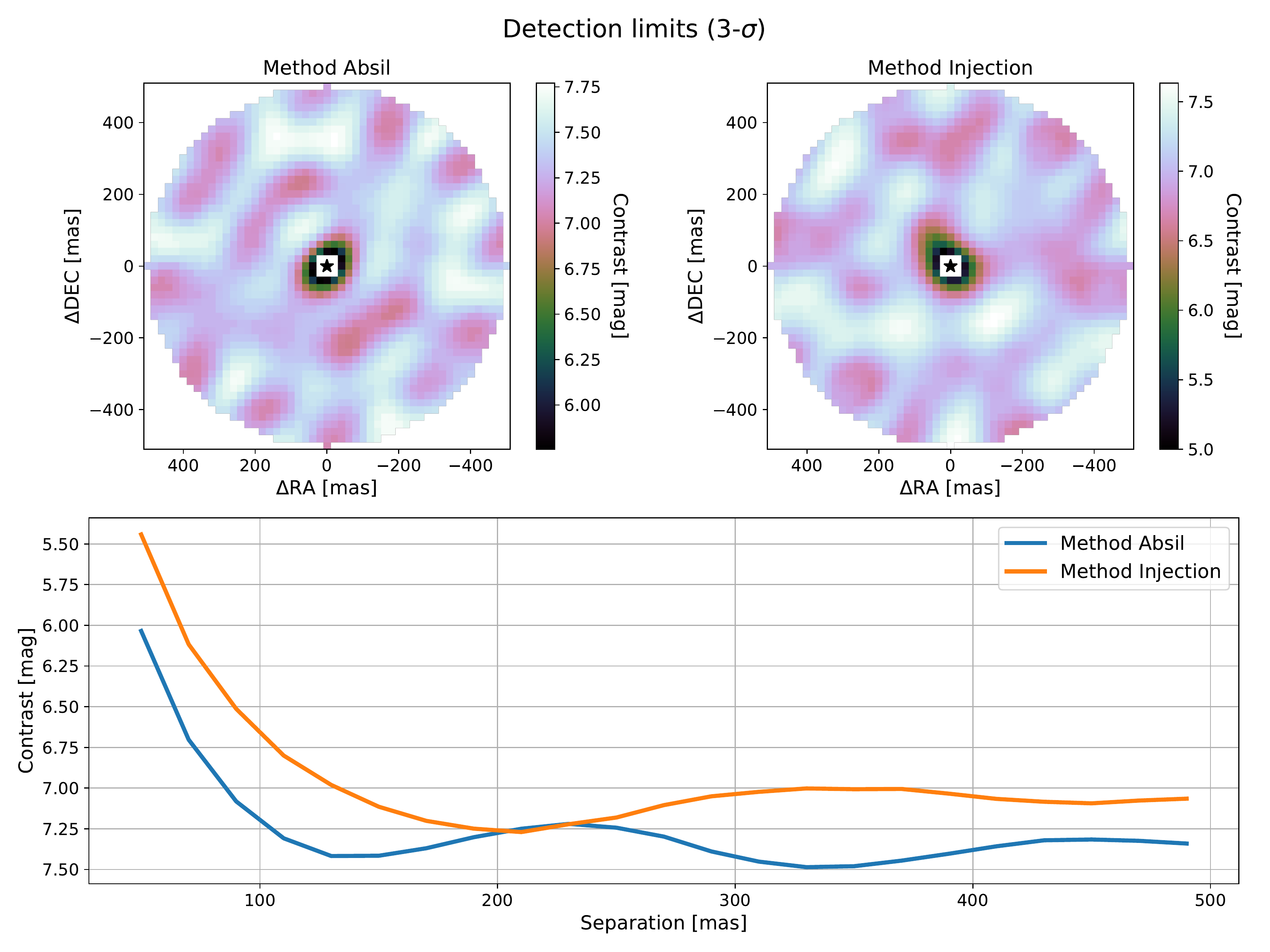}
  \caption{$3\sigma$ detection limits for AMI commissioning AB-Dor observations, after analytically removing the best-fit binary model of the known companion, AB-Dor C,  from the data. Two methods determine limits for a $10^8$ photon AMI exposure in F480M. The achievable contrast for an exposure of this depth is $\sim 7.5$ magnitudes.}
  \label{abdor_detlim}
\end{figure*}

AMI's primary commissioning program, JWST Program 1093 (PI Thatte) was designed to demonstrate binary point source detections at modest 50:1 contrasts, to exercise dithering and subdithering target placement after a target acquisition, and to determine the peak pixel value at which charge migration non-linearity remains below 1\%.  The science target AB Dor and two vetted calibrators HD~37093 and HD~36805 were observed.   Four suspected isolated K giants provided NIRISS kernel phase commissioning using AMI's observing template.  Integrations exceeding the recommended $30,000~e^-$ signal limit in the brightest pixel were used to measure charge migration. 

\subsection{AMI Science Readiness Criterion}   

The $\sim$4.5 magnitude ($\sim$50:1) contrast binary AB Dor AC was observed in the three medium-band AMI filters. Given a fully calibrated functioning instrument, the planned $\sim$10$^{8}$ photon exposure depth is more than sufficient for such a modest contrast goal.  However, this exercise was tailored to uncover unexpected major problems. The point source companion  detection limit for such exposures should normally be on the order of 7.5 magnitudes at $3\sigma$ based on observable uncertainties from \mirage~ simulations.

Pre-commissioning analysis of AB Dor \mirage~  simulations yielded confident detections of the binary companion at its predicted position and contrast within the $3\sigma$ confidence limits defined for science readiness. We used the binary fitting algorithm implemented in \texttt{fouriever} to retrieve the best-fit binary model to the data and compute a $3\sigma$ contrast curve (after analytically removing the best fit binary). Figure \ref{abdor_detlim} shows the contrast curves for simulated data using two detection limit calculation methods in \texttt{fouriever}; the injection method shows an achievable contrast $\sim$0.5 magnitude short of the $\sim$7.5 magnitude contrast photon noise floor based on Equation~\ref{eqn:contrast_limit}.

\subsection{Charge Migration}

Localized cross-talk between pixels occurs in HgCdTe detectors.  Current AMI analyses attempt to calibrate out intra-pixel capacitance (IPC) and intra-pixel sensitivity (IPS) (\eg~ \citet{2008SPIE.7021E..2BH,2014SPIE.9154E..2DH} and references therein) by using a calibrator star, rather than correct for it in early JWST pipeline processing.  Charge Migration (CM) (\eg~\citet{2018AJ....155..258C} and references therein), sometimes referred to as charge diffusion, was treated differently than IPC in commissioning data analysis in that an observation to determine its magnitude was executed.   

CM moves photoelectrons from well-illuminated pixels to less-illuminated neighbouring ones. It affects high contrast imaging of point-like sources most.  AB Dor AC (rather than one of the calibrators)  was observed,  at $\sim$50:1 contrast it is sufficiently point-like to reveal the existence of CM.  Up-the-ramp F480M exposures with peak pixel counts approaching  detector saturation (72,000$~e^-$) enabled initial linearity measurements as well as a CM estimate.  When the brightest pixel reached the AMI-specific signal limit of 30,000$~e^-$, CM stayed below 1\%.  
Summed count rates of the eight pixels surrounding the brightest pixel were compared with that of the brightest pixel as a function of integration time.  This method distinguishes CM from true pixel non-linearity. Early efforts to correct CM with the same Fourier power minimization algorithm used to correct individual bad pixels have shown promising results; work to characterize CM more fully is ongoing.
Improved CM calibration would enable higher observing efficiency for AMI and KPI since the signal limit could be increased beyond 30,000$~e^-$.

\subsection{Kernel Phase}

Four putative isolated K giants were observed with the F480M filter and AMI mode target acquisition to estimate NIRISS KPI detection limits. Targets 2MASS~J062802.01-663738.0, TYC 8906-1660-1, CPD-66 562, and CPD-67 607 brightnesses are all  $\text{W2} \sim 8~\text{mag}$.  At suitable apparent magnitudes  K giants are often isolated stars, as their age makes it unlikely that surrounding structure, if it even exists, will contribute IR excess.  Given their history, planet formation in their vicinity is exceedingly unlikely.

One of the four KPI targets (CPD-66 562) turned out to have a $\sim1$:5 contrast companion candidate at $\sim150$~mas separation easily detectable in the kernel phase observables and another $\sim1$:170 contrast detection was made around 2MASS~J062802.01-663738.0. For a detailed analysis of the KPI commissioning data see Kammerer et al. 2022 (accepted). Companion detection limits supported by this data  reached the same $\sim5$~mag contrast limit at $1~\lambda/D$, which was the AMI commissioning goal \citep[see also][for a direct comparison between AMI and KPI]{kammerer2022arXiv220800996K}.  These data supported the suggestion, based on analysis of simulated data \citep{2019JATIS...5a8001S}, that AMI should slightly outperform KPI within $\lambda/D$ for stars brighter than $M_s = 9~\text{mag}$.  


\section{Conclusion}
\label{sec:conclusions}

As the first cold space-based IR interferometer, JWST NIRISS' AMI brings unmatched image stability and photometry to the problem of detecting faint structure close to bright objects. With 3--5 \nicemicron~wavelength coverage and operating at about 40K, it provides negligible thermal infrared noise.  JWST's thermal and optical stability results in unprecedented fringe amplitude calibration, which  will encourage new image reconstruction approaches that can resolve current ambiguities in source structure, an exciting synergy with the unmatched angular resolution of today's largest ground-based telescopes.

\begin{acknowledgements}
JWST is a partnership between NASA, ESA, and CSA.  The NIRISS instrument was funded by the Canadian Space Agency and built in Canada by Honeywell.  AS and AZG acknowledge support from the NSF, NASA, and the STScI Director's Discretionary Fund, JSB acknowledges the full support from the  CONACyT ``Ciencia de Frontera'' project CF-2019/263975, 
PGT is grateful for support from the Australian Research Council grant DP1801034089, and BJSP under DE21010163.

\end{acknowledgements}

\bibliographystyle{apj.bst}
\bibliography{bibliography.bib}

\appendix{}
\section{NIRISS AMI instrument details}
\label{apdx:instrument}


The all-reflective NIRISS $f/8.7$ optical train creates an image on a HAWAII-2RG detector doped for a 5~\nicemicron ~long wavelength cutoff\footnote{\href{https://jwst-docs.stsci.edu/jwst-near-infrared-imager-and-slitless-spectrograph/niriss-instrumentation/niriss-optics-and-focal-plane}{NIRISS Optics and Focal Plane}}. Its 18~\nicemicron ~square detector pixels are formally Nyquist-spaced at $\sim$4 \nicemicron. An example simulated 80x80 AMI PSF is shown in Figure \ref{fig:pup_psf_mtf}.

\subsection{The Non-Redundant Mask}
\begin{figure*}[ht!]
    \centering
    \includegraphics[scale=.4]{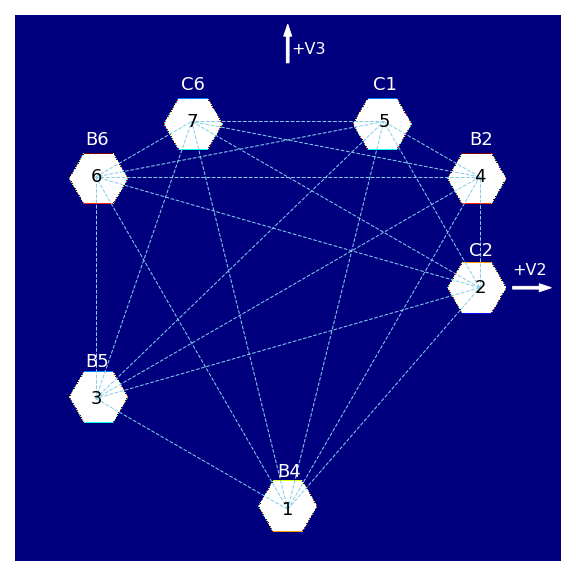}
    \caption{ NIRISS' NRM in JWST primary mirror (PM) V2-V3 coordinates after distortion-free projection back to the primary.  Seven hexagonal subpertures, 0.82~m flat-to-flat when projected to V2-V3 space, provide 21 center-to-center baselines with lengths between 1.32~m and 5.28~m. Each hole is nominally centered on a PM segment. Undersized subapertures mitigate against pupil misalignment and  shear. Hexagonal holes maximize throughput pupil  but tolerate up to 3.8\% PM diameter misalignment. Hole locations maximise long baselines while preserving non-redundancy.  The NRM exposes $\sim$15\% of the JWST PM.}
    \label{fig:nrm_baselines}
\end{figure*}

AMI on JWST is implemented by deploying a seven hole NRM in NIRISS' pupil wheel\footnote{\href{https://jwst-docs.stsci.edu/jwst-near-infrared-imager-and-slitless-spectrograph/niriss-instrumentation/niriss-pupil-and-filter-wheels}{NIRISS Filters}} (Figure \ref{fig:g7s6clearp}). 
The seven identical hexagonal subapertures are approximately 0.82~m flat-to-flat, with center-to-center baseline lengths between 1.32~m 5.28~m. Each aperture is mapped to a specific mirror segment, and the apertures are undersized compared to the mirror segments' projection to mitigate effects from edges and supporting structures. Most of the incoming light is blocked by the mask: the throughput enabled by the subapertures is $\sim$15\%. The baselines formed by this mask are described in Table~\ref{tab:baselines}.

\begin{deluxetable}{cccc}
\label{tab:baselines}
\tablecaption{21 baselines formed by 7 NRM subaperture centers projected with no distortion onto JWST's primary mirror.}
\tablehead{
\colhead{\textbf{Hole Pair}} & \colhead{\textbf{Length}} &  \colhead{\textbf{Angle}} \\
\colhead{} & \colhead{m} & \colhead{deg}
}
\startdata
        1,2 (B4--C2) & 3.492 & -40.89 \\
        1,3 (B4--B5) & 2.640 & 60.00 \\
        1,4 (B4--B2) & 4.573 & -30.00 \\
        1,5 (B4--C1) & 4.759 & -13.90 \\
        1,6 (B4--B6) & 4.573 & 30.00 \\
        1,7 (B4--C6) & 4.759 & 13.90 \\
        2,3 (C2--B5) & 4.759 & 106.10 \\
        2,4 (C2--B2) & 1.320 & 0.00 \\
        2,5 (C2--C1) & 2.286 & 30.00 \\
        2,6 (C2--B6) & 4.759 & 73.90 \\
        2,7 (C2--C6) & 3.960 & 60.00 \\
        3,4 (B5--B2) & 5.280 & -60.00 \\
        3,5 (B5--C1) & 4.759 & -46.10 \\
        3,6 (B5--B6) & 2.640 & 0.00 \\
        3,7 (B5--C6) & 3.492 & -19.11 \\
        4,5 (B2--C1) & 1.320 & 60.00 \\
        4,6 (B2--B6) & 4.573 & 90.00 \\
        4,7 (B2--C6) & 3.492 & 79.11 \\
        5,6 (C1--B6) & 3.492 & 100.89 \\
        5,7 (C1--C6) & 2.286 & 90.00 \\
        6,7 (B6--C6) & 1.320 & -60.00 \\
\enddata
\tablecomments{Angle is given in degrees counter-clockwise from the positive V3 axis. Nomenclature matches that of Figure \ref{fig:nrm_baselines}}
\end{deluxetable}

\begin{figure*}[ht!]
  \centering
  \includegraphics[scale=0.20]{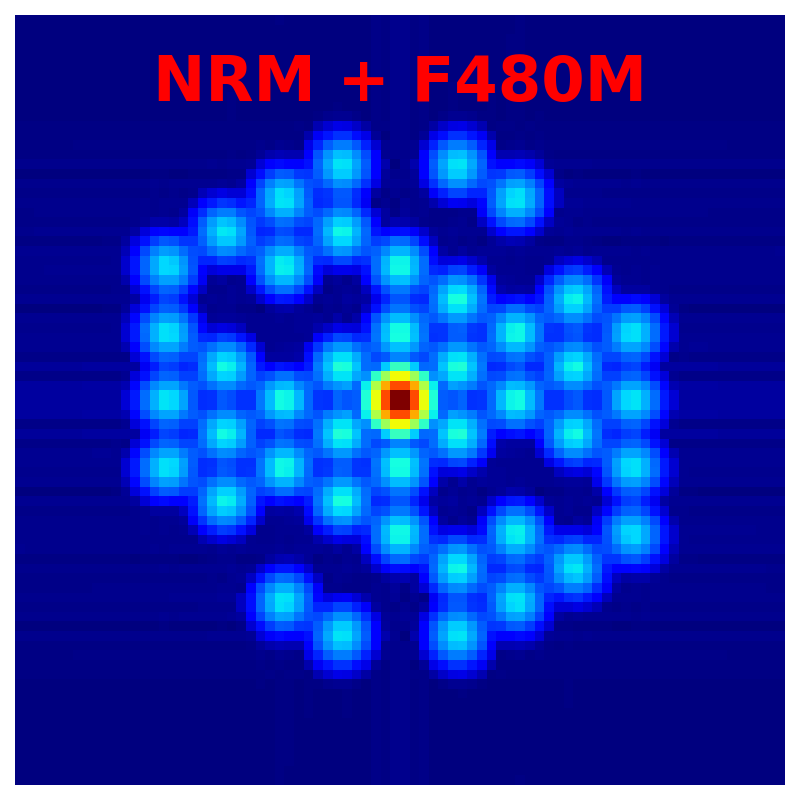} 
  \includegraphics[scale=0.20]{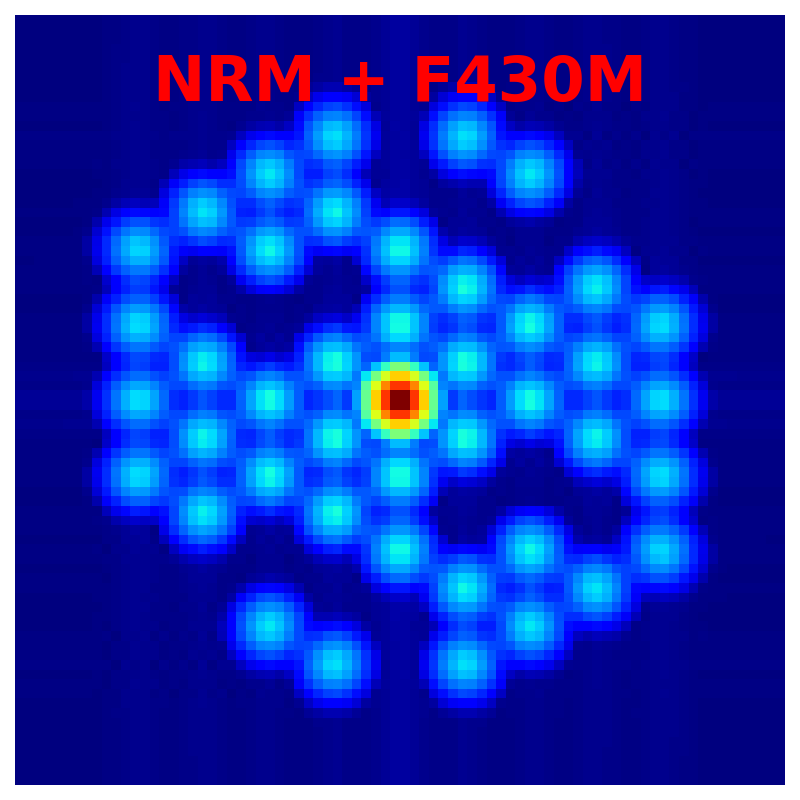}
  \includegraphics[scale=0.20]{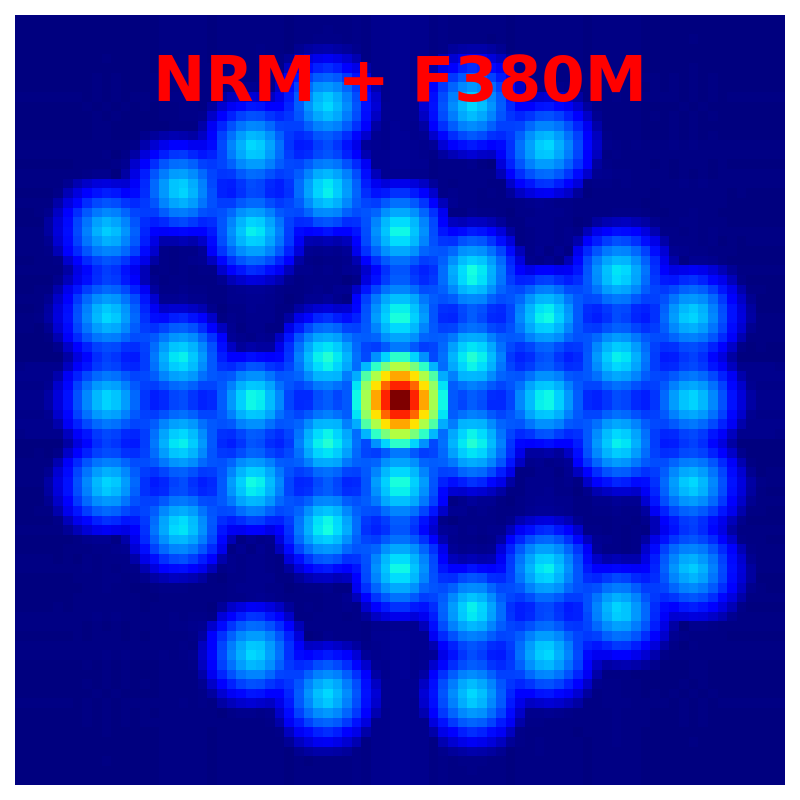}
  \includegraphics[scale=0.20]{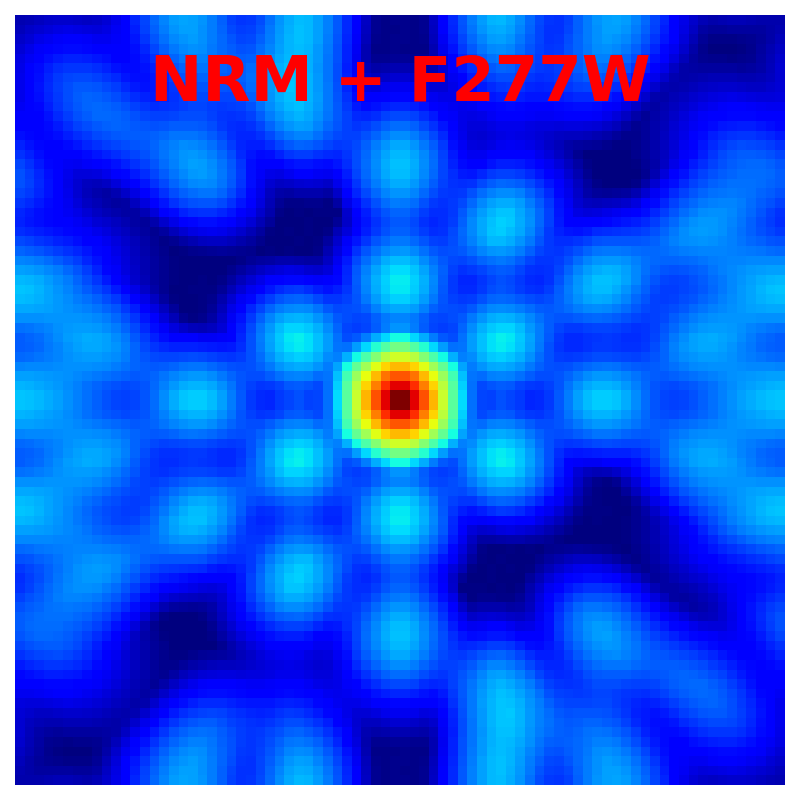} 
  \includegraphics[scale=0.20]{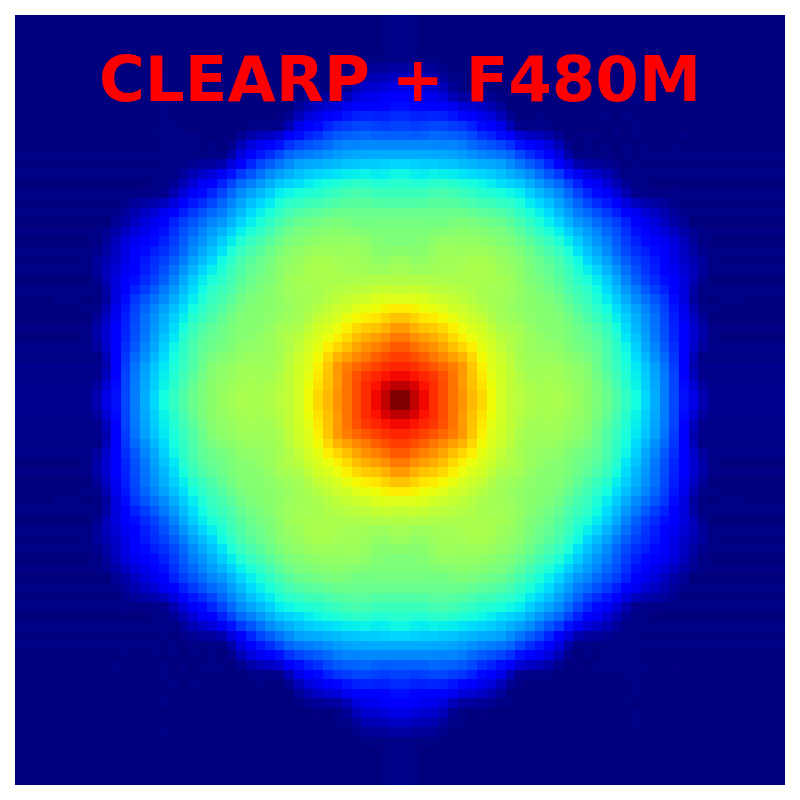} 
  \includegraphics[scale=0.20]{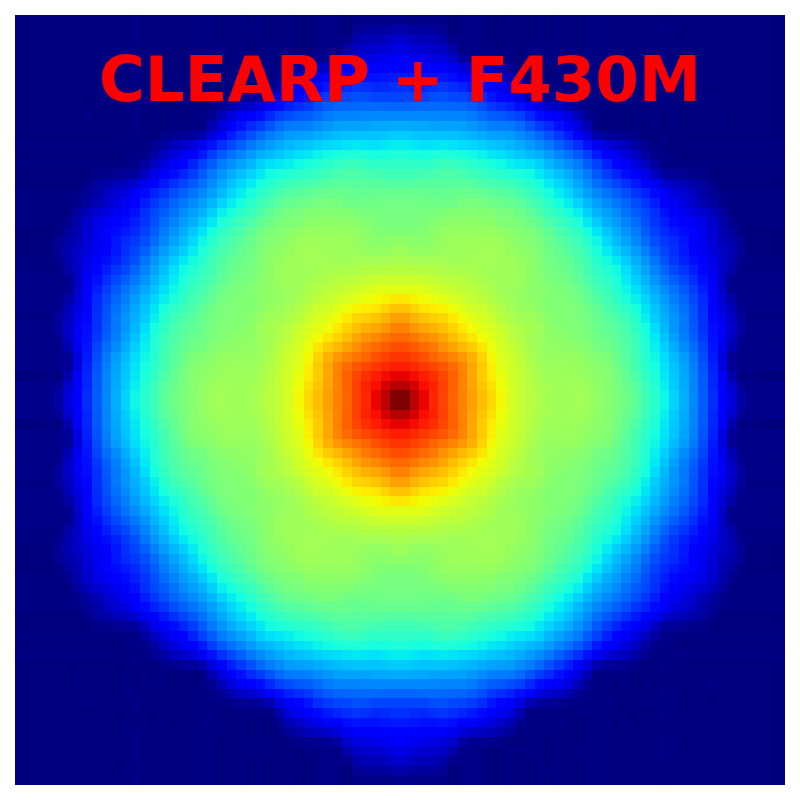}
  \includegraphics[scale=0.20]{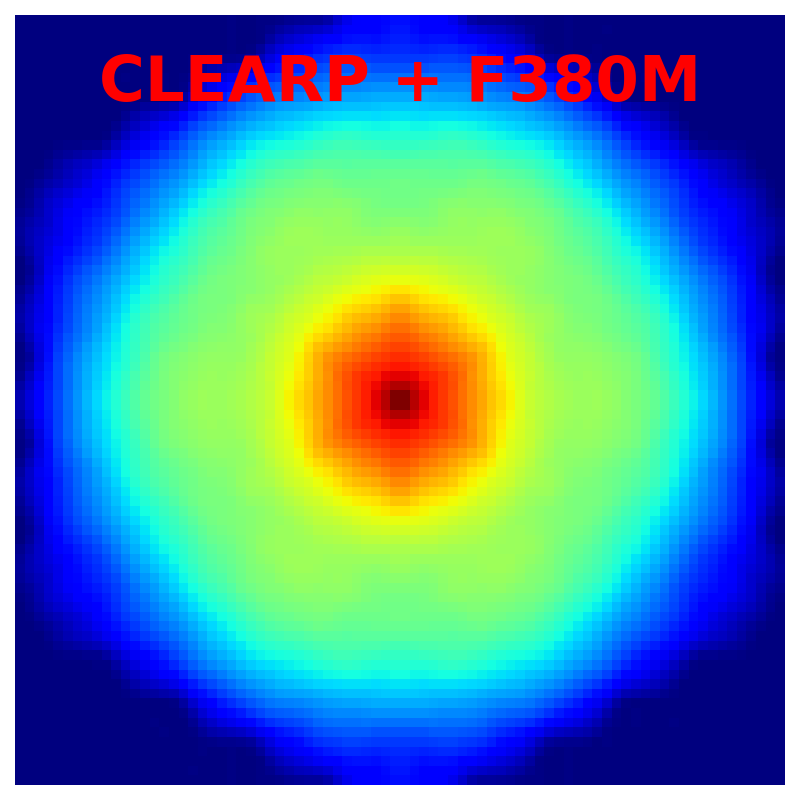}
  \includegraphics[scale=0.20]{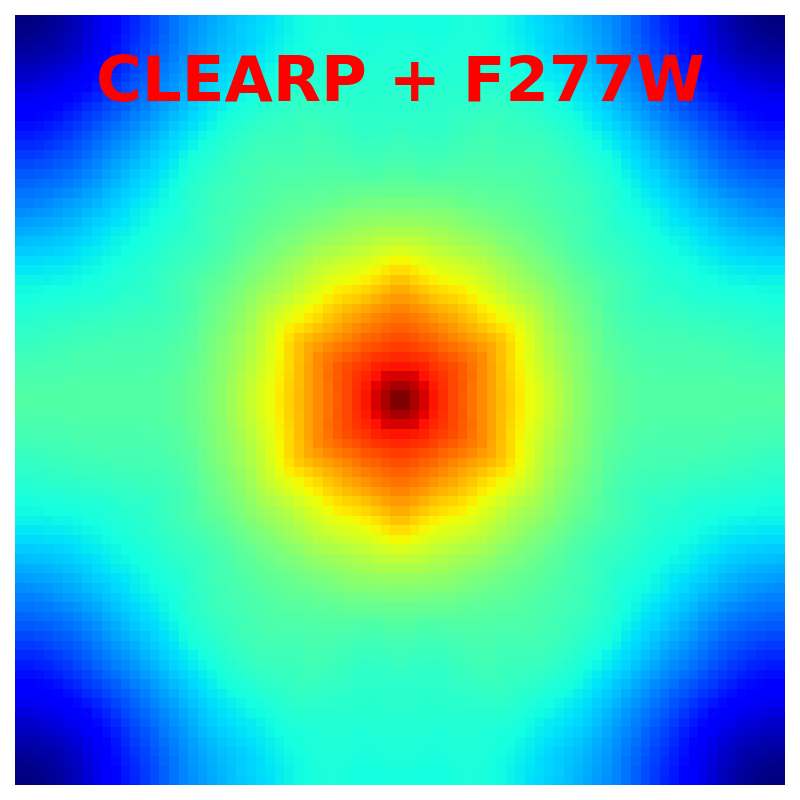}
  \caption{NRM and CLEARP \uvplane~ coverage.  AMI's pixel scale is barely Nyquist-spaced at 3.8~\nicemicron.   Fourier aliasing seen in NRM+F277W's \uvplane~  reduces AMI performance.  This may be  mitigated by understanding where aliased signal is misplaced, and treating it appropriately when possible.}
  \label{fig:filtersuv}
\end{figure*}

\subsection{AMI filters}
The NIRISS NRM can be used in conjunction with the F277W, F380M, F430M, or F480M filters. These filters were selected to cover wavelength regimes relevant to AMI's science goals. The bandpasses are relatively narrow to enable well-defined splodge locations and preserve the non-redundancy of the Fourier ($uv$) coverage. As shown in Figure \ref{fig:filtersuv}, the Fourier coverage of the three medium-band filters (F380M, F430M and F480M) comprises all baselines formed by the mask. On the other hand, the F277W broadband filter, operating at shorter wavelengths, gives access only to the central (lower spatial frequency) splodges, and there is aliasing visible at the highest spatial frequencies accessible. By combining the four filters, scientific programs using AMI will have access to photometry covering from $\sim$2.5 to $\sim$5 \nicemicron~(with gaps between the filters). More information on the NIRISS filters and recommended observing strategies is available in JWST documentation\footnote{\href{https://jwst-docs.stsci.edu/jwst-near-infrared-imager-and-slitless-spectrograph/niriss-observing-strategies/niriss-ami-recommended-strategies}{NIRISS AMI observing strategies}}.

\begin{figure*}[ht!]
    \centering
    \includegraphics[scale=.7]{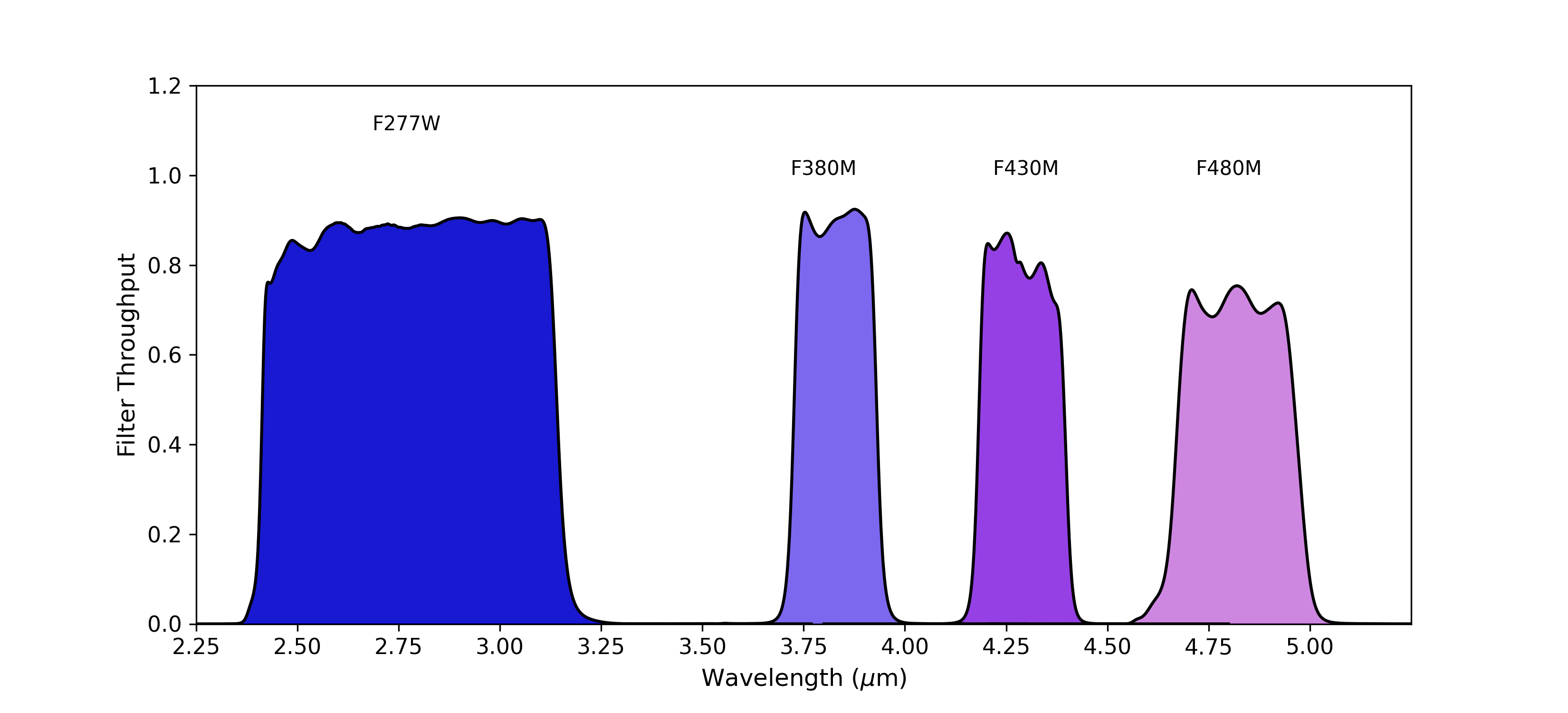}
    \caption{AMI filter throughputs.  The primary AMI filters are three medium band filters, F480M, F430M, and F380M.  They provide Nyquist or finer sampled images.  The shorter wavelength F277M filter has reduced  AMI performance but is offered because of a water band in this filter's bandpass.}
    \label{fig:ami_filters}
\end{figure*}

%
%
\begin{deluxetable}{cccc}[ht]
\tablecaption{AMI and KPI image distortion.  
 Pixel scales for 5 points within the SUB80 (or NIS\_AMI1) aperture.
 }
\tablehead{
\colhead{$x$} & \colhead{$y$} & \colhead{$p_x$ } & \colhead{$p_y$} \\
\colhead{pixels} & \colhead{pixels} & \colhead{mas/pixel} & \colhead{mas/pixel}
}
\startdata
  46.0 &  41.0 &  65.3249 &  65.7226 \\
  0.5  &   0.5 &  65.3291 &  65.7431 \\
  80.5 &   0.5 &  65.3248 &  65.7456 \\
  80.5 &  80.5 &  65.3215 &  65.7063 \\
   0.5 &  80.5 &  65.3254 &  65.7045 \\ 
\enddata

\label{tab:pixelscales}

 \tablecomments{The first row is the reference pixel position (XSciRef, YSciRef in SIAF) and the other rows are the 4 subarray corners. Data file headers contain the $2 \times 2$ linear distortion CD matrix evaluated at the reference position, in equatorial coordinates. \\
 The matrix includes a small shear as well as the differing $x$ and $y$ pixel scales. From JWST Commissioning  Program 1088, PI S. Sohn.}
\end{deluxetable}

\begin{deluxetable*}{cccccccc}
\tablecaption{Filter properties for NIRISS AMI and KPI}
\tablehead{
\colhead{Filter} & \colhead{$\lambda_{\mathrm{pivot}}$} & \colhead{$\Delta\lambda/\lambda$} & \colhead{IWA} & \colhead{CPF-AMI} & \colhead{CPF-KPI} & \colhead{AMI FOV}  & \colhead{AMI Bright limit}  \\
\colhead{} & \colhead{\nicemicron} & \colhead{\%} & \colhead{mas} & \colhead{}    & \colhead{}   & \colhead{arcsec}       & \colhead{ Vega mag } 
} 
\startdata
F277W & 2.78 & 26.3 & 89 & 0.045$^*$ & 0.23$^*$ & 0.79 & 7.5 \\
F380M & 3.83 & 5.4 & 120 & 0.026     & 0.13$^*$ & 1.15 & 4.6 \\
F430M & 4.29 & 5.0 & 140 & 0.022     & 0.11$^*$ & 1.28 & 4.0 \\
F480M & 4.82 & 6.3 & 150 & 0.018     & 0.08     & 1.44 & 3.6 \\
\enddata

\tablecomments{ 
    $\lambda_{\mathrm{pivot}}$ is defined in \citet{2005PASP..117..421T}. \newline
    $\Delta\lambda$: half-power filter width. \newline
    IWA: Inner Working Angle. \newline
    CPF: Central Pixel Fraction.  CPFs from simulations (starred) may be $\sim$10\% high. \newline
    AMI FOV: interferometric field of view,  \ie, the primary beam core's diameter. \newline
    KPI's bright limit is $\sim$4.5 magnitudes fainter than AMI's.
} 
\tablerefs{Values found in updated \href{https://jwst-docs.stsci.edu/about-jdox}{JWST documentation} should be used if they differ from the above. Early results are presented here.}
\end{deluxetable*}

\end{document}